\documentclass[%
 aip,
 amsmath,amssymb,
 reprint,%
]{revtex4-1}

\usepackage{graphicx}
\usepackage{dcolumn}
\usepackage{bm}

\usepackage[utf8]{inputenc}
\usepackage[T1]{fontenc}
\usepackage{mathptmx}
\usepackage{etoolbox}

\makeatletter
\def\@email#1#2{%
 \endgroup
 \patchcmd{\titleblock@produce}
  {\frontmatter@RRAPformat}
  {\frontmatter@RRAPformat{\produce@RRAP{*#1\href{mailto:#2}{#2}}}\frontmatter@RRAPformat}
  {}{}
}%
\makeatother
\begin{document}

\preprint{AIP/123-QED}

\title[A Cryogenic Uniaxial Strain Cell for Elastoresistance Measurements]{A Cryogenic Uniaxial Strain Cell for Elastoresistance Measurements}
\author{F. Eckelt}
\email{eckelt@uni-wuppertal.de}
\affiliation{\mbox{University of Wuppertal, School of Mathematics and Natural Sciences, 42097 Wuppertal, Germany}}
\affiliation{Wuppertal Center for Smart Materials, University of Wuppertal, Germany}

\author{I. Cavallaro}
\affiliation{\mbox{University of Wuppertal, School of Mathematics and Natural Sciences, 42097 Wuppertal, Germany}}

\author{R. Wagner}
\affiliation{\mbox{University of Wuppertal, School of Mathematics and Natural Sciences, 42097 Wuppertal, Germany}}

\author{A.-A. Haghighirad}
\affiliation{\mbox{Karlsruhe Institute of Technology, Institute for Quantum Materials and Technologies, 76021 Karlsruhe, Germany}} 

\author{T. Wolf}
\affiliation{\mbox{Karlsruhe Institute of Technology, Institute for Quantum Materials and Technologies, 76021 Karlsruhe, Germany}} 

\author{J. Wilmers}
\affiliation{\mbox{Stralsund University of Applied Sciences, Faculty of Mechanical Engineering, 18435 Stralsund, Germany}}
\affiliation{Chair of Solid Mechanics, School of Mechanical Engineering and Safety Engineering, University of Wuppertal, Germany}
\affiliation{Wuppertal Center for Smart Materials, University of Wuppertal, Germany}

\author{S. Bargmann}
\affiliation{Chair of Solid Mechanics, School of Mechanical Engineering and Safety Engineering, University of Wuppertal, Germany}
\affiliation{Wuppertal Center for Smart Materials, University of Wuppertal, Germany}

\author{C. Hess}
\affiliation{\mbox{University of Wuppertal, School of Mathematics and Natural Sciences, 42097 Wuppertal, Germany}}
\affiliation{Wuppertal Center for Smart Materials, University of Wuppertal, Germany}

\date{\today}

\begin{abstract}
We present the design, implementation, and validation of a cryo-compatible strain cell for elastoresistance measurements in quantum materials. The cell is actuated by three large-format piezoelectric stacks and enables both compressive and tensile strain up to approximately $\pm5\%$. The relative displacement of the sample holders is measured in situ using a high-resolution capacitive displacement sensor, ensuring precise strain control throughout measurement.
To operate the strain cell over a broad temperature range from 4.2\,K to 300\,K, a modular measurement probe was developed, allowing efficient thermal coupling and integration into cryogenic environments.
The functionality and precision of the setup were demonstrated by elastoresistance measurements on the iron-based superconductor BaFe$_2$As$_2$ along the [110] direction. The results were validated against conventional techniques based on glued samples and strain gauges, showing excellent agreement in amplitude and temperature dependence of the elastoresistance coefficient. Additional tests revealed that the system remains robust under mechanical vibrations introduced by vacuum pumps, enabling stable operation even with liquid nitrogen cooling. This highlights the high mechanical and thermal reliability of the developed platform for low-temperature transport measurements under controlled uniaxial strain.
\end{abstract}

\maketitle

\section{\label{sec:level1} Introduction}

Uniaxial pressure represents an effective tuning parameter for selectively modulating the electronic properties of materials. In contrast to hydrostatic pressure, uniaxial pressure can directly break the rotational symmetry of a crystal, often leading to distinct effects on electronic behavior. Representative systems include Sr$_2$RuO$_4$ and YBa$_2$Cu$_3$O$_{6.67}$. While hydrostatic pressure causes a gradual reduction of the superconducting transition temperature in Sr$_2$RuO$_4$~\cite{Dehnungszelle11}, uniaxial pressure leads to a rapid enhancement of the same~\cite{Dehnungszelle21,Dehnungszelle3}. In contrast, hydrostatic pressure enhances superconductivity in YBa$_2$Cu$_3$O$_{6.67}$~\cite{Dehnungszelle4}, whereas uniaxial pressure along the \textit{a}-axis suppresses superconductivity and can simultaneously induce static charge order~\cite{Dehnungszelle5}.\\
The simplest method for applying uniaxial stress involves the use of mechanical springs and screw mechanisms, which has been realized in various setups~\cite{MechanischStrain1,MechanischStrain2,MechanischStrain3,Chu2010}. The main drawback of this approach lies in the limited ability to tune the strain \textit{in situ}.
Another method employs gas-filled bellows, which allow for \textit{in situ} control of the applied strain by pushing the sample holders apart~\cite{BlasBalck}. Alternatively, bending devices can be used, in which the desired strain is introduced through flexure of the sample substrate~\cite{Biegen11,Biegen21}.\\
Gluing samples directly onto piezoelectric actuators enables compact designs with fully \textit{in situ} adjustable strain. However, this method also transfers both the lateral expansion of the actuator and its thermal expansion to the sample. Moreover, the maximum achievable strain is limited by the mechanical properties of the actuator itself. 
An elegant and mechanically stable alternative is a strain cell consisting of three piezoelectric actuators. In this design, the piezoelectric actuators displace a sample holder, allowing the full actuator stroke to be used for strain, which enables the application of significantly larger uniaxial strains \cite{Dehnungszelle1}.
In this contribution, we extend this principle by developing a version of the cell with a larger form factor, which accommodates more powerful actuators capable of sustaining higher forces. It also provides increased space for sample mounting and features a modular construction that facilitates experimental flexibility and customization. In addition to the strain cell itself, a dedicated measurement probe was developed to enable its use in a cryostat at low temperatures, and a measurement program was implemented to perform elastoresistance measurements.

\section{\label{sec:level1}Measurement probe}

\begin{figure}
\includegraphics[width=0.25\textwidth]{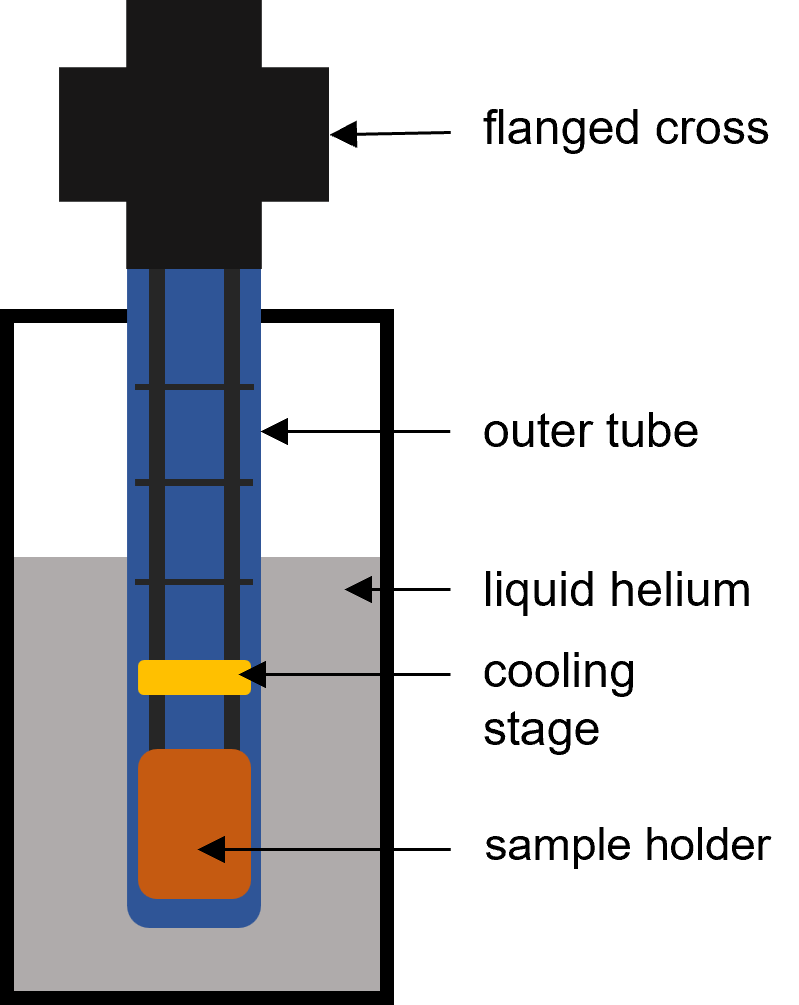}
\caption{\label{fig:Design_Experiment} Schematic setup of the experiment. The developed strain cell is mounted onto the sample holder of the custom-designed measurement probe, which is enclosed within a stainless steel tube and can be evacuated via the flange head. Thermal contact to the cryogen is established through a cold stage.}
\end{figure}

\begin{figure*}[]
\centering
\includegraphics[width=0.8\textwidth]{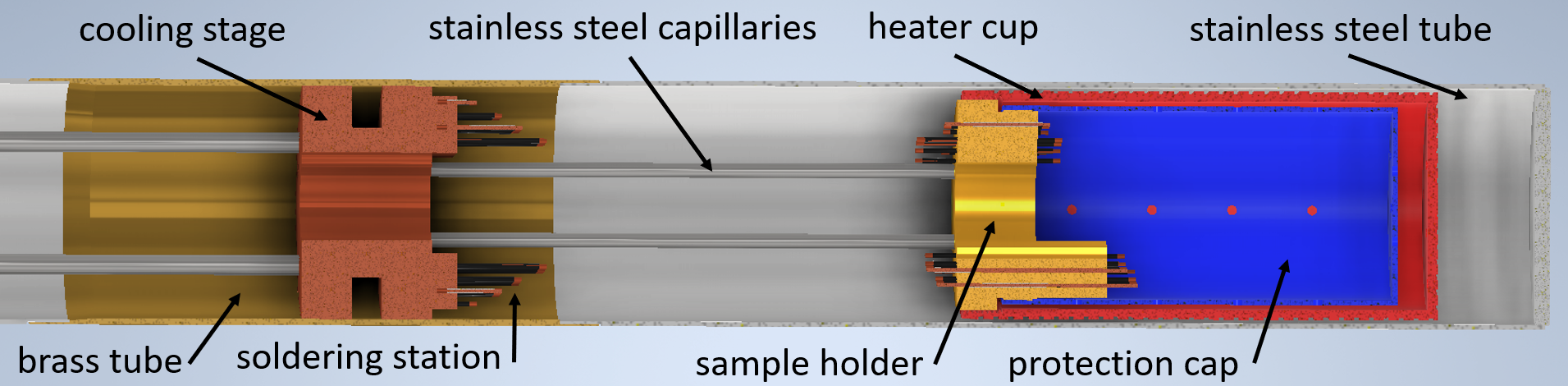}
\caption{\label{fig:Aufbau_Kaelte}Lower end of the probe.  
At the bottom of the probe, the sample holder is mounted to four stainless steel capillaries, which are connected to the probe’s flange head. For temperature control, a protective cap and a heater cup made of gold-plated copper can be screwed onto the sample holder. A total of 28 electrical connections are available at the sample holder, eight of which are used for temperature measurements using a Cernox 1050-SD-HT and a Pt100 sensor. Thermal contact to the helium bath is established via a copper cold stage, which is thermally anchored to the outer stainless steel tube through a soldered-in brass sleeve. To minimize thermal load from the measurement wires, they are thermally anchored at the cold stage via a soldering platform and routed in pairs through fabric sleeves inside the capillaries for mechanical and thermal protection.}
\end{figure*}

\begin{figure*}[]
\centering
\includegraphics[width=0.8\textwidth]{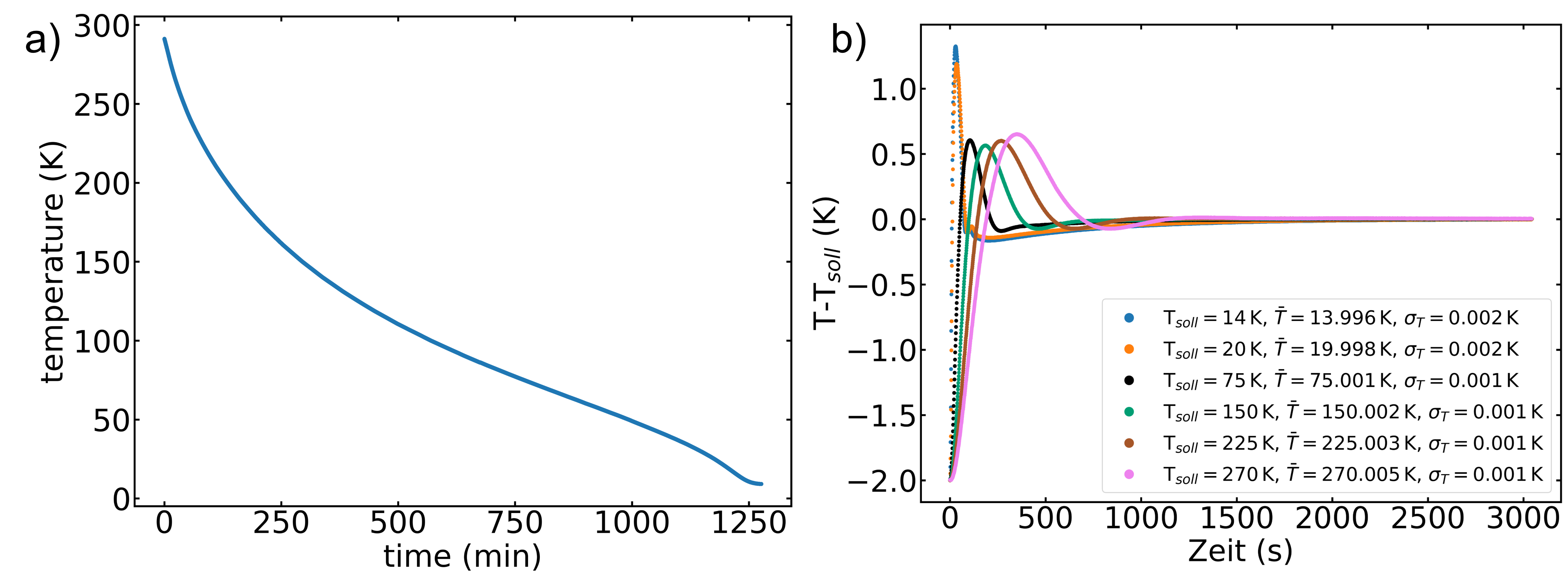}
\caption{\label{fig:Temperaturregelung}a) Cooldown curve of the probe using liquid helium. The base temperature of 9\,K is reached after approximately 1250\,minutes. The cooldown process can be significantly accelerated by introducing helium exchange gas into the interior of the probe. 
b) Temperature regulation of the sample for various target temperatures. To allow for better comparison, the difference between the actual sample temperature and the setpoint is plotted. The legend includes the mean temperature ($\bar{T}$) and standard deviation ($\sigma_T$) in the time interval between 2000\,s and 3000\,s after the setpoint change.}
\end{figure*}

The developed strain cell was designed to be integrated into a custom-built probe assembly. This probe was intentionally constructed as a flexible platform suitable for a variety of low-temperature experiments. It enables precise temperature control in the range from 4.2\,K to 300\,K. For this purpose, it can be inserted into a cryostat or a transport dewar filled with liquid helium, as shown in Figure~\ref{fig:Design_Experiment}. The probe is based on a tube design, in which the sample holder is mounted to a flange head via four stainless steel capillaries and inserted into a stainless steel tube with an outer diameter of 49.5\,mm. The interior of the probe can be evacuated through the flange head. Thermal contact between the sample holder and the helium bath is established via a cold stage, which is shown in detail in Figure~\ref{fig:Aufbau_Kaelte}. The cold stage consists of a copper block mounted between the stainless steel capillaries. At this position, a brass tube is soldered into the outer tube. Due to the higher thermal expansion coefficient of brass compared to copper, the brass tube contracts more strongly during cooldown, creating tight mechanical contact with the copper block \cite{martienssen2005springer}. This enables efficient thermal coupling between the cold stage and the helium bath. Using this setup, the sample holder can be cooled down to 9\,K within approximately 1250\,minutes (Figure~\ref{fig:Temperaturregelung}a)). By introducing helium exchange gas into the interior of the probe, the cooling process is significantly accelerated and the base temperature reduced to approximately 4.2\,K, reducing the total cooling time to just a few hours.

The sample holder is equipped with a total of 28 electrical contacts. Two groups of four contacts each are dedicated to four-point resistance measurements using a Cernox 1050-SD-HT sensor (Quantum Design, USA) and a Pt100 sensor (Testo SE \& Co. KGaA, Germany) for temperature monitoring. Electrical connections are made using $50\,$µm thick copper wires running from the sample holder to the flange head. To minimize thermal load on the sample, these wires are thermally anchored to the cold stage via a soldering platform. From there, they continue to the flange head, routed through fabric sleeves inside the stainless steel capillaries for mechanical stabilization. For ease of use in four-point resistance measurements, the wires are grouped in sets of four and connected to vacuum feedthroughs at the top of the probe, which likewise support four-wire connections. To reduce electromagnetic interference, the measurement lines are twisted in pairs. Temperature control is achieved via a heater cup made of gold-plated copper, which can be mounted onto the sample holder. The outer surface of the heater contains a double spiral into which two parallel-connected, $100\,$µm thick manganin wires are glued. These wires have a total length of approximately 2\,m and a combined resistance of $170\,\Omega$. To shield the sample from direct thermal radiation emitted by the heater, an additional protective cap made of gold-plated copper is installed over the sample holder (Figure~\ref{fig:Temperaturregelung}a)). The screw-on thermal radiation shield defines a usable mounting area of 34\,mm\,$\times$\,90\,mm (width\,×\,length). The maximum usable height is limited by the circular cross-section and reaches up to approximately 25\,mm, depending on the width of the sample.
A Lakeshore 340 temperature controller (Lake Shore Cryotronics) is used for temperature regulation. It provides a maximum output power of 100\,W at up to 50\,V and 2\,A.
The power is regulated via an internal PID controller that adjusts the output voltage. Given the heater resistance of $170\,\Omega$, the maximum achievable heating power is approximately 14\,W.  

Figure~\ref{fig:Temperaturregelung}b) shows the temperature regulation performance for various setpoints. For better comparison, the difference between the measured sample temperature and the target temperature is plotted over time. The data demonstrate that the target temperature can be precisely reached across the entire temperature range. A small overshoot is observed at the beginning of each regulation cycle, with its magnitude slightly depending on the setpoint. To evaluate the stability, the temperature mean value $\bar{T}$ and standard deviation $\sigma_T$ are indicated in the legend for the interval between 2000\,s and 3000\,s after the setpoint change. The largest deviation of the mean temperature from the setpoint was only 5\,mK at 270\,K. The highest observed standard deviation was just 2\,mK at 14\,K and 20\,K, highlighting the high quality of the temperature control system.

\section{\label{sec:level1}Strain cell}

\begin{figure*}[]
\centering
\includegraphics[width=0.65\textwidth]{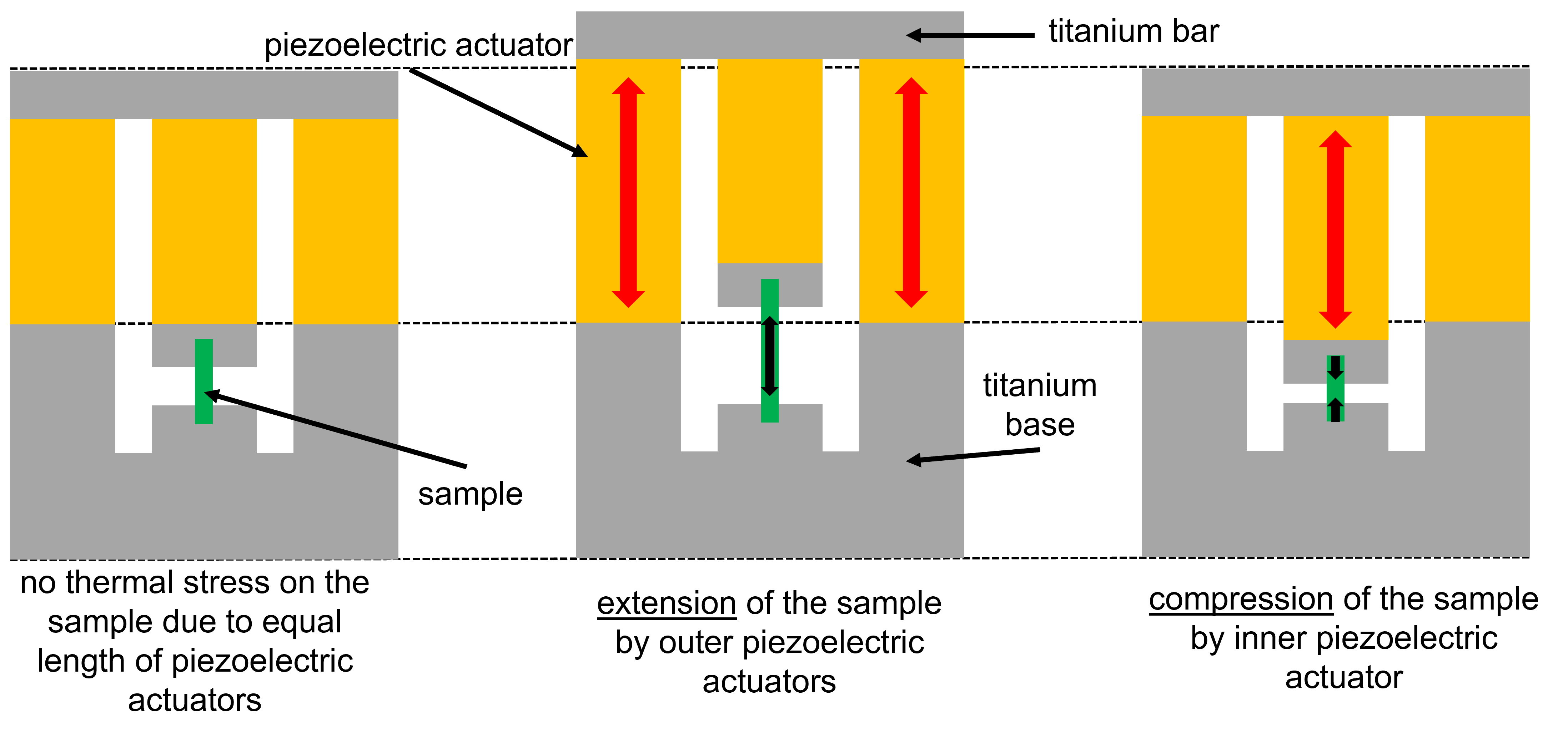}
\caption{\label{fig:Skizze} Operating principle of the strain cell. The use of three piezoelectric actuators of equal length ensures compensation of thermal expansion, allowing for strain-free cooling of the sample. Tensile strain is applied by extending the two outer actuators (center), while compressive strain is induced by extending the central actuator (right).}
\end{figure*}

The strain cell is based on a concept by Hicks et al. ~\cite{Dehnungszelle1} and essentially consists of three piezoelectric actuators that generate the mechanical load on the sample. The setup and working principle of this approach are illustrated in Figure~\ref{fig:Skizze}. The sample (green) is mounted between two sample holders, which can be displaced relative to each other by means of the three piezo actuators (orange). When the two outer actuators are extended, a tensile strain is applied to the sample (center). Conversely, extending the central actuator compresses the sample (right). Since all three piezo actuators are of equal length, their thermal expansion compensates during cooldown, allowing the sample to reach cryogenic temperatures without mechanical stress.

\begin{figure}[]
\centering
\includegraphics[width=0.45\textwidth]{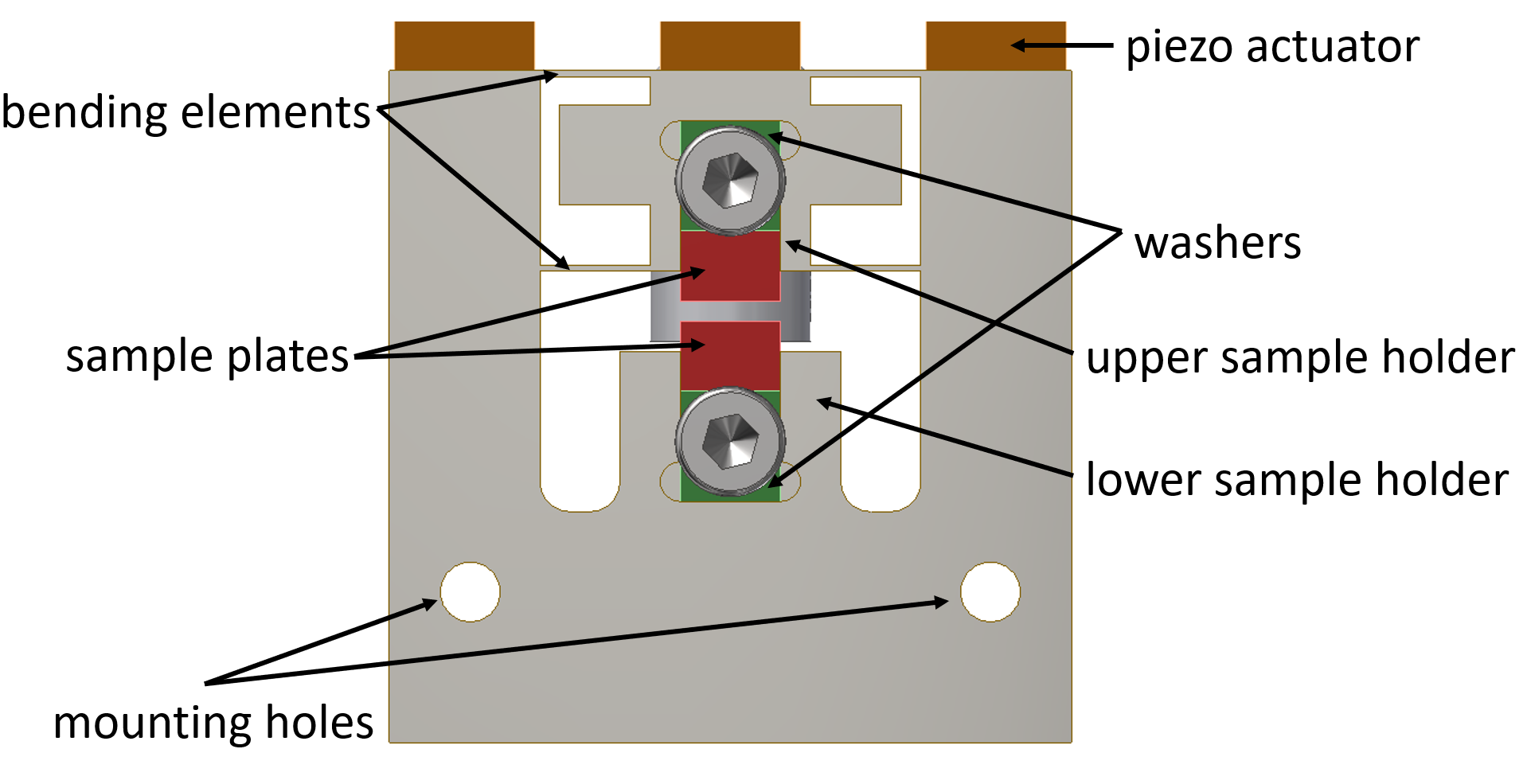}
\caption{\label{fig:Draufsicht_Titanunterteil}Top view of the titanium base section of the strain cell.}
\end{figure}

Figure~\ref{fig:Draufsicht_Titanunterteil} shows a top view of the lower part of the strain cell. The sample plates are used to mount both ends of the sample. The distance between these plates determines the maximum achievable strain according to $\epsilon = \Delta L / L_0$, where $\Delta L$ is the displacement generated by the piezo actuators and $L_0$ is the initial distance between the sample plates in the unloaded state. By varying the length of the plates, $L_0$ and thus the maximum achievable strain can be adjusted. The maximum distance between the sample holders is 4\,mm. The strain cell can be mounted to the probe via the mounting holes.

To stabilize the motion of the upper sample holder against rotational and lateral displacement, and to protect the central piezo actuator from unintended lateral forces, for example during sample mounting, it is connected to the titanium base via four bending elements. Due to their geometry, these elements exhibit a low spring constant for axial motion, while providing significantly higher stiffness against rotational or transverse movements.

\subsection{\label{sec:level2}Finite element analysis}

\begin{figure}[]
\centering
\includegraphics[width=0.45\textwidth]{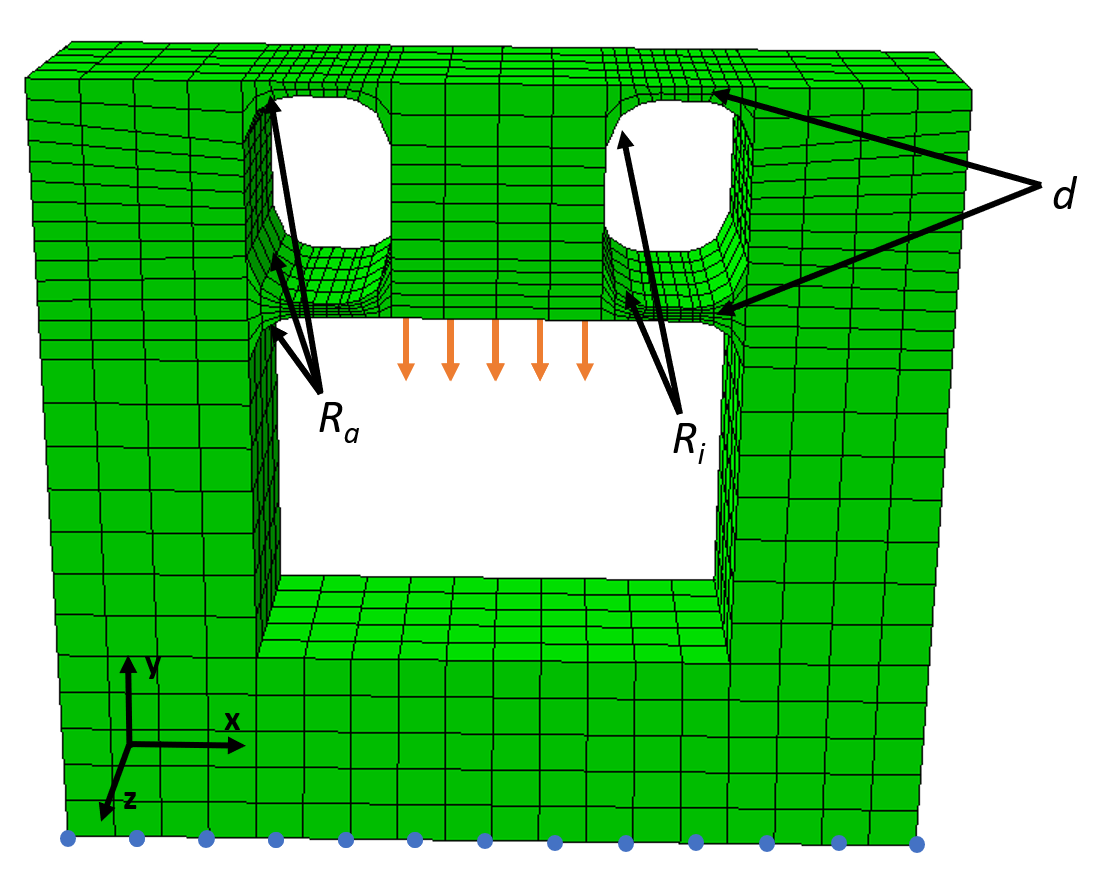}
\caption{\label{fig:Konfiguration} Structured mesh applied to the entire geometry of the component. The mesh is locally refined in critical areas, especially where the bending elements connect to the top of the cell, to ensure greater accuracy in areas of expected higher stress or strain. The bottom of the component is fixed for the simulation, as indicated by the blue points. The force is applied to the upper sample holder, as shown by the orange arrows.}
\end{figure}

To verify the mechanical behavior of the bending elements and determine the thickness $d$ and radii $R_i$ and $R_a$ at the junctions with the rest of the strain cell (see Figure~\ref{fig:Konfiguration}), a finite element analysis was conducted using Abaqus CAE. A mesh composed of hexahedral elements with 8 nodes, reduced integration (C3D8R) and hourglassing control was employed. This choice of elements provides a good balance between computational effort and accuracy, particularly for three-dimensional structures. The material, assumed to be linear elastic, was titanium grade 2 with a Young's modulus of 103 MPa and a Poisson's ratio of 0.33 \cite{Titan}.\\

A structured mesh was used (see Figure~\ref{fig:Konfiguration}). In regions of particular interest, where higher stresses or strains were expected, the mesh was locally refined. This specifically applies to the upper part of the cell, where the bending elements are connected to the rest of the component. This refinement ensures that the stress distribution in these critical areas is accurately captured.
To accurately replicate the function of the cell in the simulation, the load on the component was applied by displacing the upper sample holder, as depicted in Figure~\ref{fig:Konfiguration}. The orange arrows illustrate the applied displacement direction. A displacement of $50\,$µm was used for the calculations. The lower part of the component was fixed, indicated by the blue points. 

\begin{figure}[]
\centering
\includegraphics[width=0.45\textwidth]{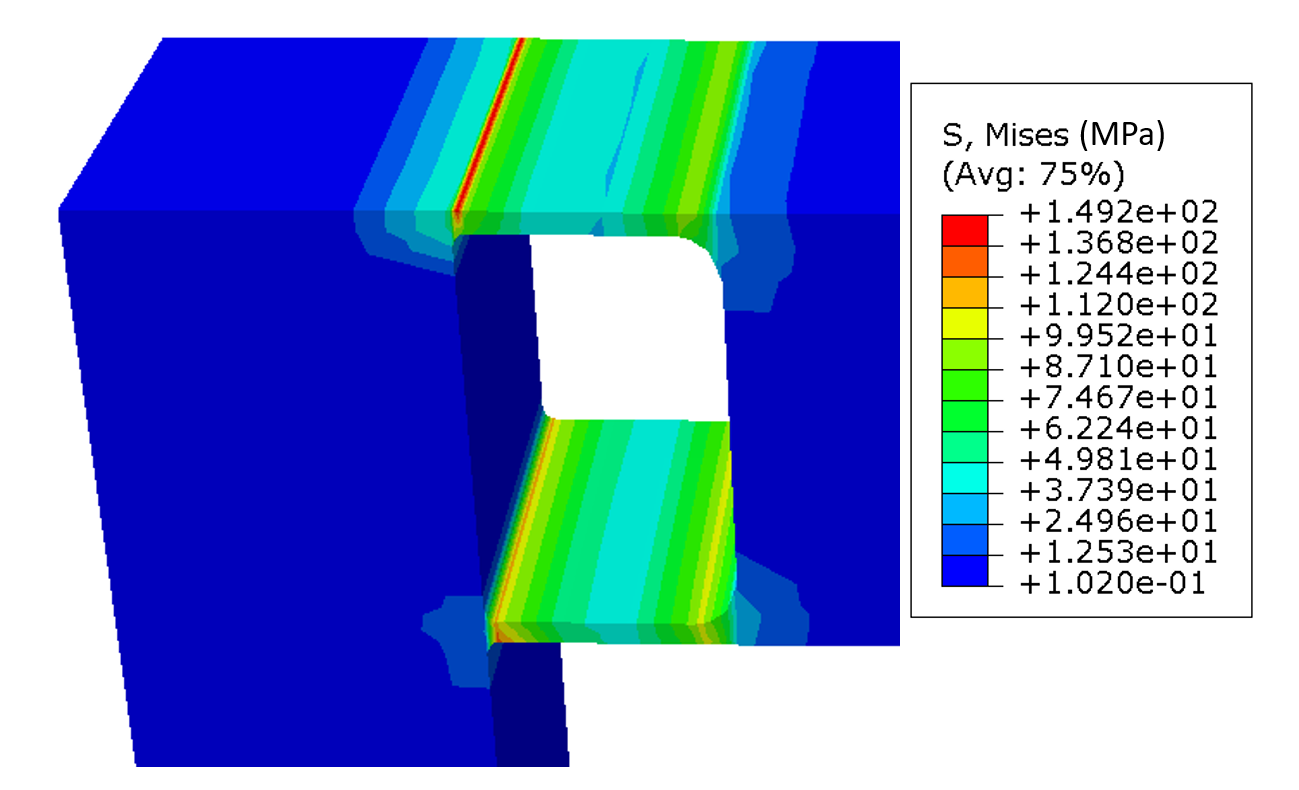}
\caption{\label{fig:MiesesSpannung} Color-coded visualization of the von Mises stress distribution on the component. The maximum stress occurs in the radii of the bending elements.}  
\end{figure}

Figure~\ref{fig:MiesesSpannung} shows a representative von Mises stress map for an imposed displacement of 50\,µm  applied to the upper sample holder. It clearly illustrates that the transition regions with radii are subject to particularly high mechanical loads. This increased stress in the transition areas is the reason why both the radii and the thickness were systematically varied in the simulations.\\
To assess the load within the component, the von Mises stress was utilized. This stress measurement combines the effects of normal and shear stresses into a single quantity, providing a conservative estimate of material failure. The calculated von Mises equivalent stress was compared to the yield strength of titanium grade 2, which is $275\,$MPa \cite{Titan}.

\begin{figure*}[]
\centering
\includegraphics[width=0.95\textwidth]{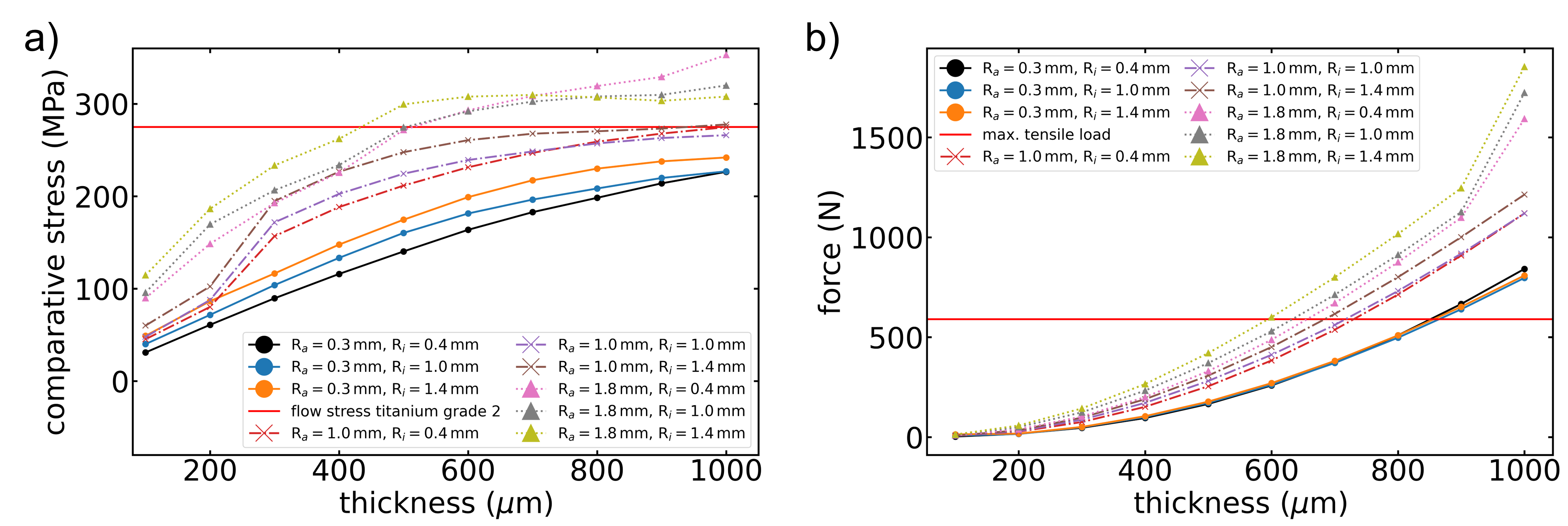}
\caption{\label{fig:ErgebnisseSimulation1} a) Simulation results showing the von Mises stress for various combinations of inner and outer radii $R_i$ and $R_a$ plotted as a function of the thickness $d$. The maximum allowable stress for titanium grade 2 is indicated in red. b) Required force for a displacement as a function of thickness $d$, inner radius $R_i$, and outer radius $R_a$. The maximum tensile force that a piezo stack can withstand is indicated in red.}
\end{figure*}

In Figure~\ref{fig:ErgebnisseSimulation1}a, the simulation results for different geometries are depicted, showing various combinations of inner and outer radii plotted versus the thickness~$d$. The output stress value is the maximum von Mises stress from all elements. Both larger inner and outer radii $R_i$ and $R_a$, as well as greater thicknesses $d$, lead to higher stresses in the component. The yield strength of titanium grade 2, which is 275\,MPa, is marked in red for comaprison. For most configurations, the stresses in the component do not reach the yield strength, indicating that the bending elements deform only elastically. \\
In Figure~\ref{fig:ErgebnisseSimulation1}b, the force required to displace the upper sample holder is shown for the same geometric configurations as in Figure~\ref{fig:ErgebnisseSimulation1}a. The plotted values represent the force required to achieve a displacement of $50\,$µm for various combinations of inner and outer radii $R_i$ and $R_a$, as a function of thickness $d$. The required force increases slightly with increasing inner and outer radii and shows a quadratic dependence on thickness.

The piezoelectric actuators used in this setup are made of sintered ceramic layers and are therefore sensitive to tensile stress. During both tensile and compressive loading of the sample, at least one actuator is always subjected to tensile force. In a uniaxial tensile test performed on one of these actuators, mechanical failure occurred at a load of 590\,N. To ensure a sufficient safety margin, the maximum operational load should remain well below this value. For comparison, this maximum load is indicated by the red line in Figure~\ref{fig:ErgebnisseSimulation1}b. Here, it should be noted that the applied load during the experiment must also cover the force required to deform the specimen itself.
The simulation results show that small transition radii $R_i$ and $R_a$ are desirable, as they reduce both mechanical stress in the component and the required actuation force. In practice, the minimum achievable radii are limited by manufacturing constraints. Using wire electrical discharge machining, a minimum radius of $0.3\,$mm can be achieved, which is sufficient to allow for a suitable component configuration. For $R_i = R_a = 0.3\,$mm, the von Mises stresses remain within the elastic range for all simulated thicknesses (see Figure~\ref{fig:ErgebnisseSimulation1}a). The required force also remains within the performance limits of the actuators for thicknesses up to $900\,$µm (see Figure~\ref{fig:ErgebnisseSimulation1}b).
A thickness of $400\,$µm was chosen to achieve a balance between high force reserves and minimal stress on the piezo stacks, as confirmed by simulation. It should be noted that during operation of the strain cell, at least one piezoelectric actuator is always under tensile load, which can reduce its lifetime and, in the worst case, lead to failure. However, due to the modular design of the strain cell, the actuators can easily be replaced and should therefore be regarded as consumable components.

\subsection{\label{sec:level4} Strain measurement}

\begin{figure}[]
\centering
\includegraphics[width=0.45\textwidth]{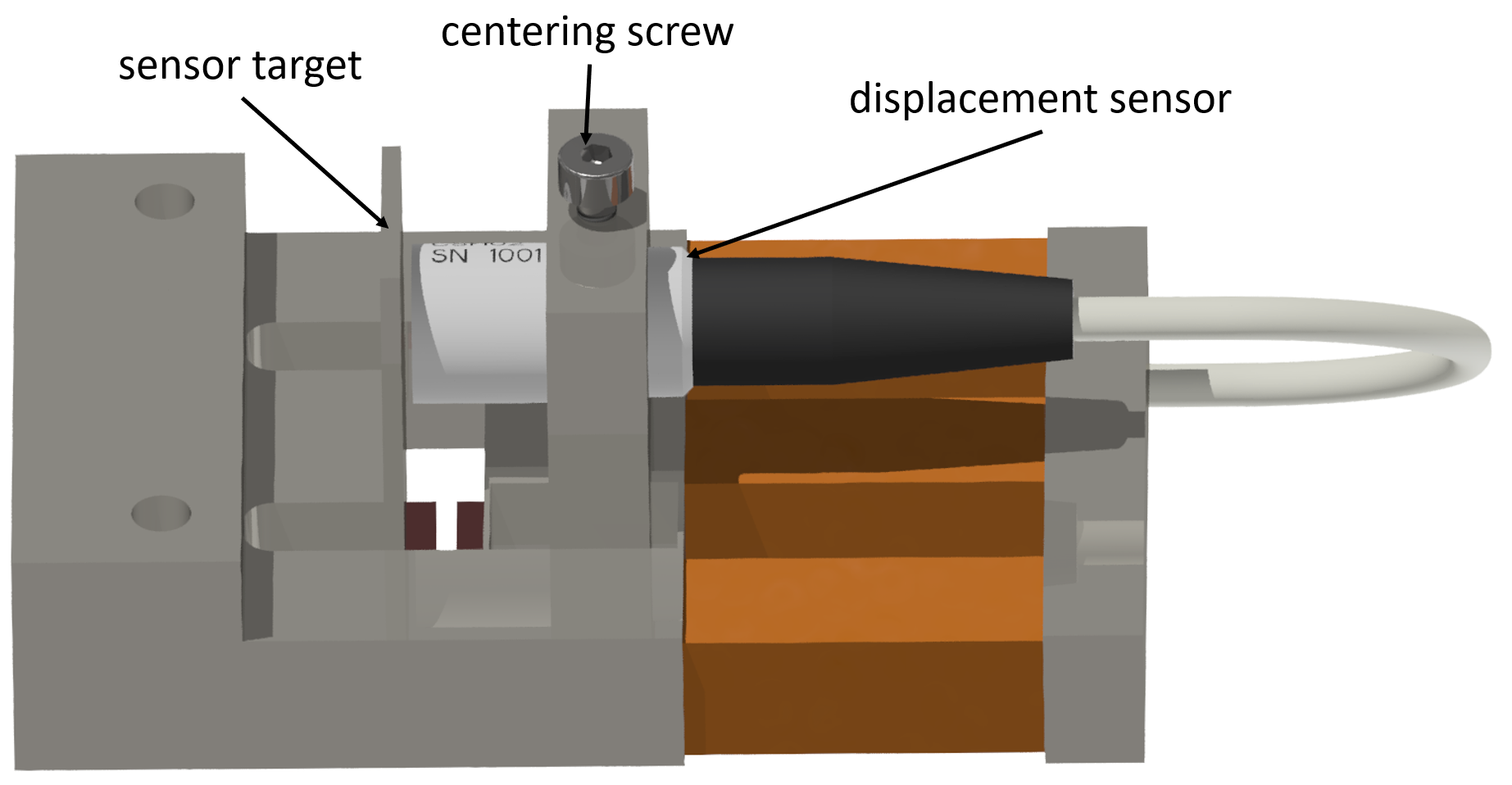}
\caption{\label{fig:UntereSeite} Lower part of the strain cell with the capacitive displacement sensor, which can be mounted to the upper sample holder using a centering screw. The sensor measures the distance to the lower sample holder via a sensor target.}  
\end{figure}

The distance between the sample holders is monitored with a capacitive displacement sensor (model CSH05; Micro-Epsilon, Germany). The sensor provides a measurement range of 1\,mm and allows for a maximum temporal resolution of up to 8.5\,kHz, at which it achieves a resolution of 10\,nm \cite{Abstandssensor}. The operating principle is based on an ideal parallel-plate capacitor, where the sensor and an opposing target form the two capacitor plates. The sensor is mounted in a holder on the underside of the strain cell and is attached to the upper sample holder, while the target is a titanium reflector mounted to the lower sample holder. For accurate operation, the target must be electrically grounded to ensure a stable electric potential and to shield the measurement signal from electrical noise and static charge accumulation. The underside of the strain cell, including labeled components, is shown in Figure~\ref{fig:UntereSeite}.

\subsection{\label{sec:level6} Influence of vacuum pumps}

\begin{figure*}[]
\centering
\includegraphics[width=0.9\textwidth]{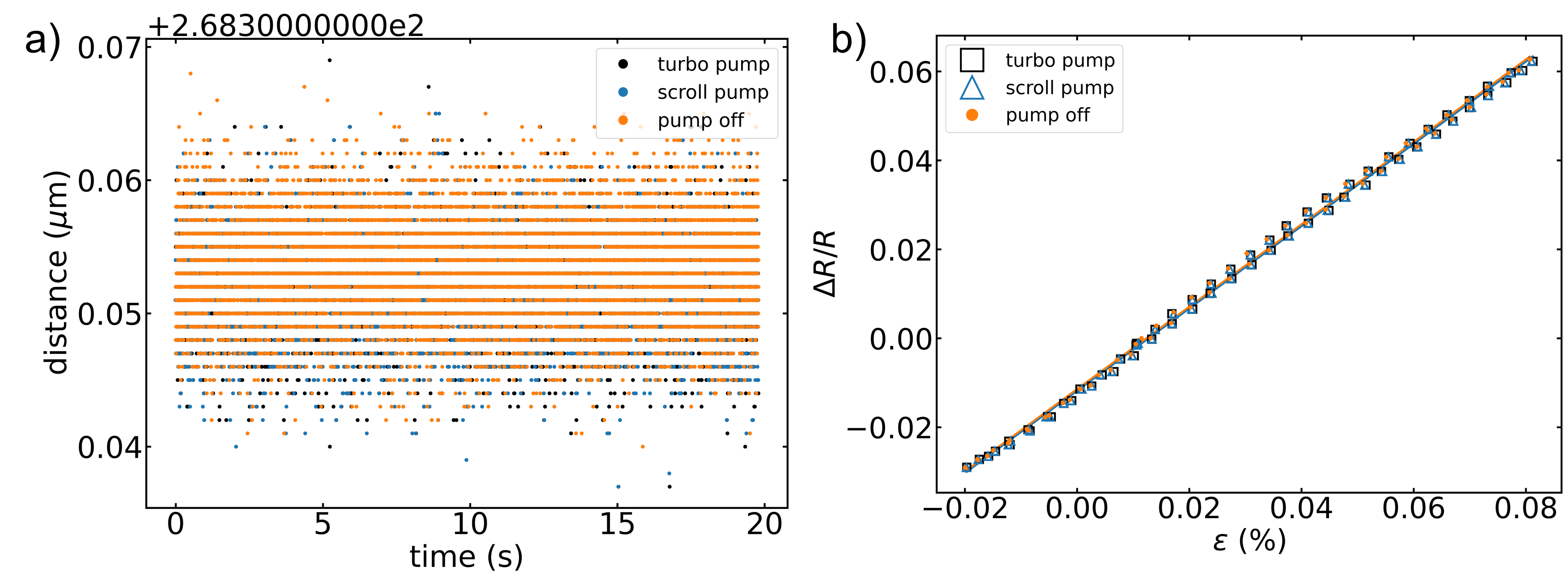}
\caption{\label{fig:Pumpeanundaus} Investigation of the influence of vacuum pumps on the measurement. a) Time-resolved measurement od the gap between the sample holders at a fixed set-point distance for different pump configurations (none, scroll pump, turbomolecular pump) show no significant effect on the signal. b) Strain-dependent resistance change measured at 138\,K on a test sample of BaFe\(_2\)As\(_2\) under the same pump configurations, also showing no measurable influence from the pumps. The individual data points are barely visible, as they lie almost entirely on top of each other.}  
\end{figure*}

During measurements in liquid nitrogen, the reduced cryopumping effect compared to liquid helium leads to a noticeable degradation of the vacuum inside the probe despite the low leak rate of the setup ($1.55 \cdot 10^{-5}\,\mathrm{mbar\cdot l/s}$). As a result, thermal coupling is enhanced and the available heating power of 14\,W is no longer sufficient to raise the sample temperature to room temperature. This issue can be resolved by evacuating the probe with a vacuum pump. To investigate the influence of vacuum pumping on the measurement, both a scroll pump and a turbomolecular pump were tested. The pumps were connected to the probe via a flexible metal hose. Figure~\ref{fig:Pumpeanundaus}a shows the 
distance measurement under different pumping configurations, revealing no significant impact on the displacement signal. In addition, measurements of the strain-dependent change in electrical resistance were carried out on BaFe$_2$As$_2$ at 138\,K (see Figure~\ref{fig:Pumpeanundaus}b) using the same pump configurations, as shown in Figure~\ref{fig:Pumpeanundaus}a. Again, no substantial influence on the data was observed. These results demonstrate that the probe can reliably be used with liquid nitrogen for extended elastoresistance measurements, without compromising the ability to heat the sample to high temperatures. Furthermore, they highlight the robustness of the developed strain cell against minor vibrations transmitted through the vacuum hose from the pumps.

\subsection{\label{sec:level5} Maximum displacement}

\begin{figure}[]
\centering
\includegraphics[width=0.45\textwidth]{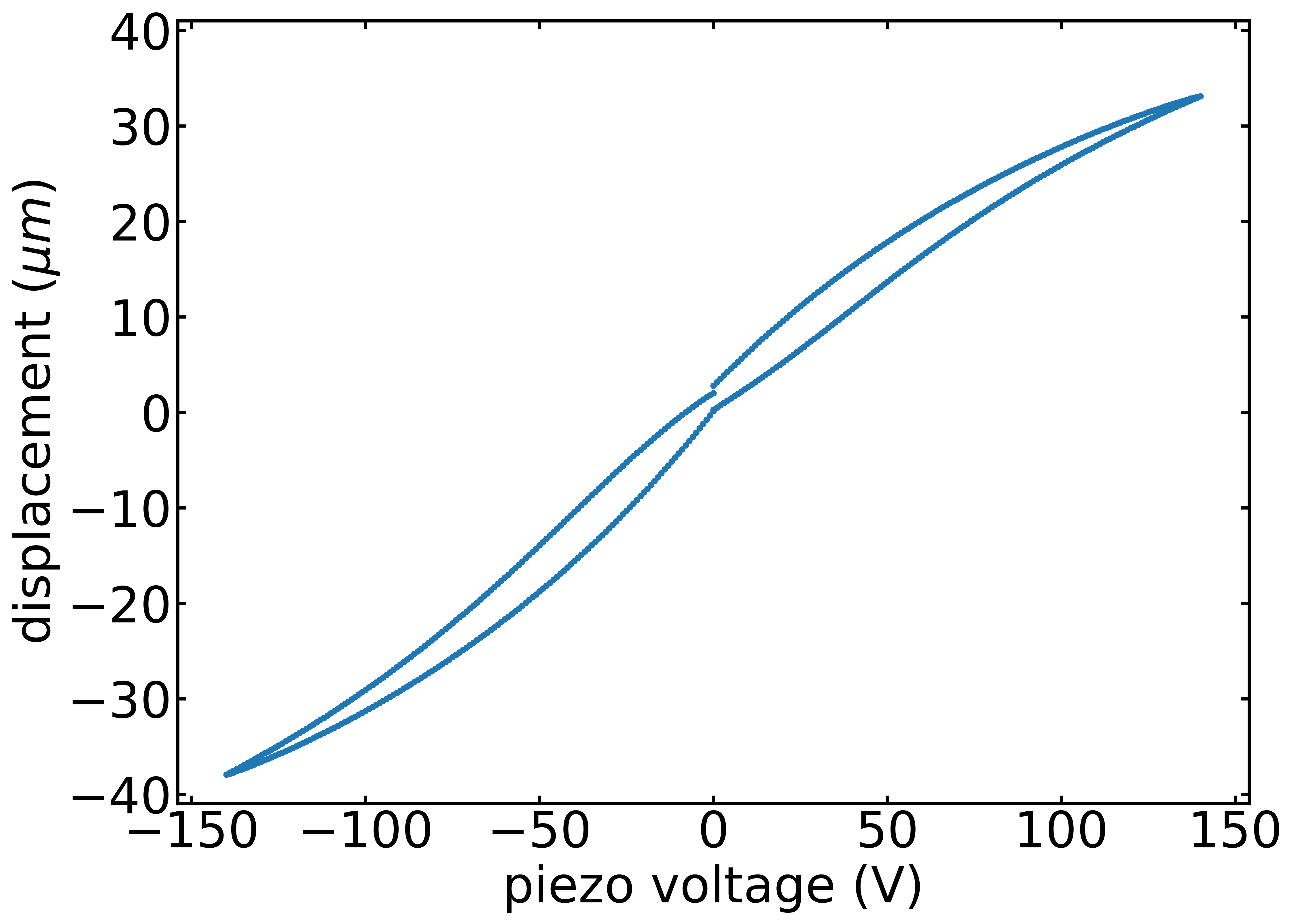}
\caption{\label{fig:Verschiebung} Displacement of the sample holders as a function of the applied piezo voltage at room temperature.}  
\end{figure}

In the final version of the strain cell, three piezoelectric actuators (model PST 150/7 × 7/50; Piezomechanik GmbH, Germany), each 45 mm long, were installed. This results in overall dimensions of the strain cell of 22\,mm (height) × 83.5\,mm (length) × 34\,mm (width). To determine the maximum displacement at room temperature, the distance between the sample holders was measured as a function of the applied piezo voltage. The result is shown in Figure~\ref{fig:Verschiebung}. The maximum displacement corresponding to tensile strain was 33\,µm, while the maximum compressive displacement reached 37\,µm. For a minimum effective sample length of 600\,µm, this corresponds to a maximum tensile strain of 5.5\,\% and a maximum compressive strain of 6\,\%.

\subsection{\label{sec:level6} Measurement software}
\begin{figure*}[]
\centering
\includegraphics[width=0.9\textwidth]{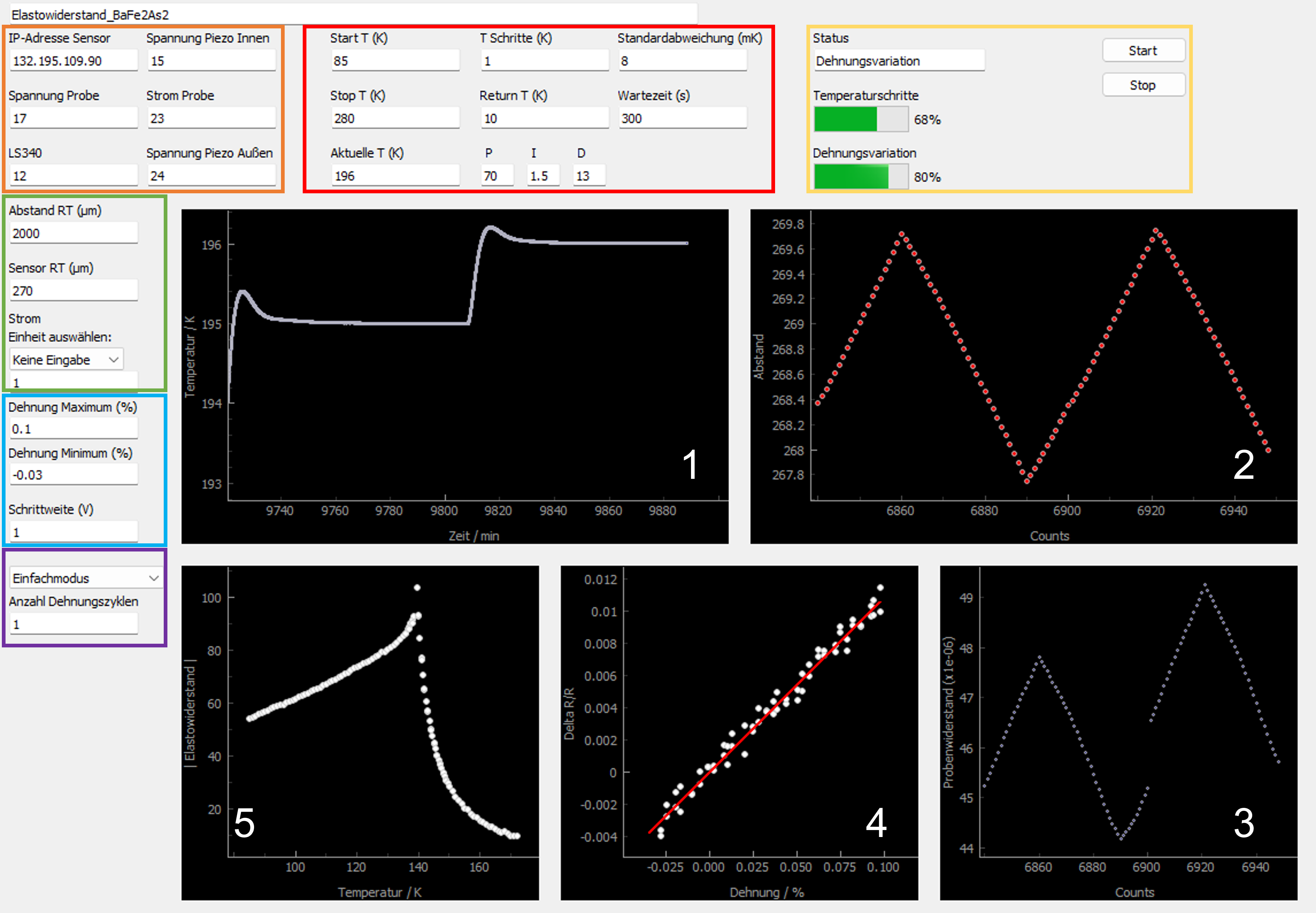}
\caption{\label{fig:Programm} Graphical interface of the developed measurement software, which enables control of the measurement electronics and real-time monitoring of the acquired data.}  
\end{figure*}

To perform elastoresistance measurements using the developed strain cell, a user-friendly software with a graphical interface was created. The program handles communication with the measurement instruments and provides real-time data visualization. The goal of elastoresistance measurements is to determine the strain-dependent change in electrical resistance at constant temperature. Figure~\ref{fig:Programm} shows the graphical user interface of the program, where the acquired data can be monitored across five plots. Plot 1 continuously displays the sample temperature, which must remain constant during the measurement. Plots 2 and 3 show the displacement and resistance as a function of the applied piezo voltage, updated in real time. In Plot 4, the relative resistance change $\,\Delta R/R_0 = (R - R_0)/R_0\,$ is plotted against the sample strain for a constant temperature, where $R_0$ denotes the resistance in the unstrained state. A linear fit is applied, and the slope corresponds to the elastoresistance coefficient. These coefficients, obtained for different temperatures, are summarized in Plot 5.

The various input fields allow the user to define all relevant measurement parameters. Communication with the measurement instruments is handled via GPIB, except for the displacement sensor, which is controlled via Ethernet. The corresponding GPIB addresses can be entered in the orange-framed fields. The red-framed section allows configuration of the temperature control parameters, including the PID values of the controller, the start and stop temperatures, and the temperature step size. In the green-framed fields, the initial distances at room temperature must be entered to enable accurate strain calculation. This section also includes the input current used for the four-point resistance measurement.The blue-framed section is used to define the target strain and the step size for the piezo voltage. In the purple-framed area, the number of strain cycles to be performed per temperature point can be specified.

\subsection{\label{sec:level6} Test measurements}
\begin{figure*}[]
\centering
\includegraphics[width=0.9\textwidth]{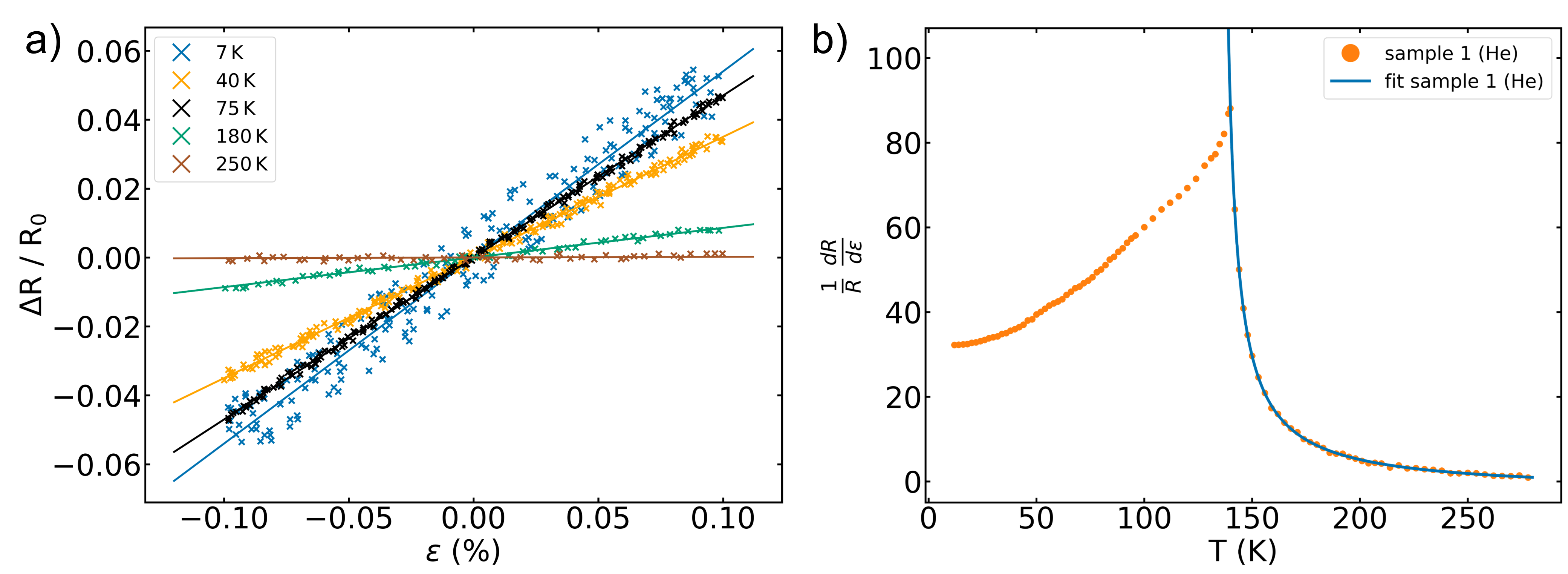}
\caption{\label{fig:Widerstandsänderung_Elasto} (a) Strain-dependent relative resistance change at various temperatures. (b) Temperature dependence of the elastoresistance coefficient with applied Curie-Weiss fit. Measurements were performed using liquid helium as the cooling medium.}  
\end{figure*}

To validate the functionality of the developed strain cell, elastoresistance measurements were performed on a single-crystalline BaFe$_2$As$_2$ sample along the [110] direction and compared to results obtained using the conventional method, in which the sample is glued directly onto a piezoelectric actuator and the strain is monitored using a surface-mounted strain gauge~\cite{Xiaochen2,Xiaochen1}. In the absence of external strain, BaFe$_2$As$_2$ undergoes an antiferromagnetic phase transition at approximately 138\,K, which is accompanied by a structural distortion that breaks the fourfold symmetry of the high-temperature tetragonal phase~\cite{Fernandes2014,Fernandes2013,test}. This orthorhombic distortion is relatively small and is driven by an electronically induced nematic instability. The electronic origin of the symmetry breaking manifests itself in a pronounced in-plane resistivity anisotropy. The strong coupling between the electronic anisotropy and the lattice makes BaFe$_2$As$_2$ an ideal test material for strain experiments. In particular, the [110] direction is highly sensitive to uniaxial strain, as it selectively couples to the nematic order parameter~\cite{Chu,Chualt,Kuo1}. \\
For the measurement, a piece of the crystal was cut into a rectangular shape along the [110] direction, resulting in final sample dimensions of 3\,mm\,$\times$\,0.5\,mm\,$\times$\,50\,µm. The distance between the sample plates was set to 1.5\,mm, and the sample was glued using Stycast 2850FT. Electrical contacts were made using four 50\,µm thick silver wires attached with Hans Wolbring conductive silver 200N. For each measurement, the sample temperature was stabilized and the strain was varied continuously between $-0.1\,\%$ and $+0.1\,\%$, while recording the resistance in real time. Figure~\ref{fig:Widerstandsänderung_Elasto}a shows the strain-dependent relative resistance change $(R - R_0)/R_0$, where $R_0$ is the resistance of the unstrained sample. At all temperatures, a linear relationship between resistance change and strain is observed. At low temperatures, the noise level increases slightly due to the lower sample resistance. The elastoresistance coefficient $\tilde{n}$ is defined as the slope of the linear dependence of the relative resistance change on strain. The temperature dependence of $\tilde{n}$ is shown in Figure~\ref{fig:Widerstandsänderung_Elasto}b and exhibits the expected increase upon cooling, followed by a decrease above the nematic transition temperature $T_{\mathrm{nem}}$. Above $T = 139$\,K, the data can be described by a Curie-Weiss-like behavior of the form
\begin{equation}
\tilde{n}=n_0+\frac{\lambda/a_0}{T-T_{\text{nem}}}
\end{equation}

as indicated by the green line in Figure~\ref{fig:Widerstandsänderung_Elasto}b ~\cite{Kuo2}.

\begin{figure*}[]
\centering
\includegraphics[width=0.9\textwidth]{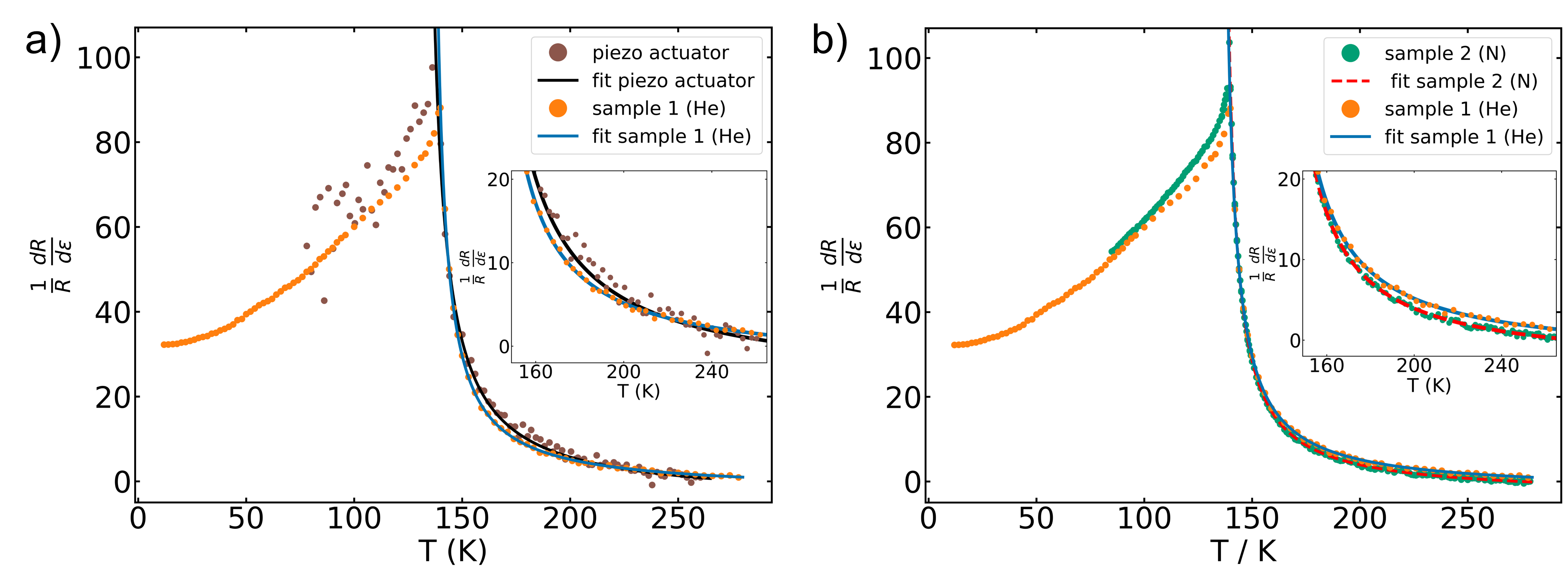}
\caption{\label{fig:VergleichElasto} Comparison of the elastoresistance measurement of sample 1 with a) an additional measurement performed using the conventional method, where the sample is directly glued onto the surface of the piezoelectric actuator, and b) a second elastoresistance measurement using the strain cell, this time employing liquid nitrogen as the cooling medium. In both panels, the temperature range between 150\,K and 260\,K is shown enlarged in an inset to facilitate comparison.}  
\end{figure*}

To confirm the measurement results, an additional elastoresistance experiment on BaFe$_2$As$_2$ was performed using the conventional method, in which the sample is directly glued onto a piezoelectric actuator. The result is shown in Figure~\ref{fig:VergleichElasto}a, together with the previously presented measurement using the developed strain cell. In this reference measurement, the expected divergent temperature dependence is again observed and can also be described by a Curie-Weiss-like behavior. A high degree of agreement is found between the two data sets. In particular, the matching amplitude of the elastoresistance signals confirms that the strain in the developed strain cell is correctly determined. Notably, the measurement using the strain cell exhibits significantly lower noise compared to the conventional method. This may be attributed to less effective strain transfer in the conventional setup.

To further assess the reliability of the measurement approach using the strain cell, a second elastoresistance experiment was conducted, this time using liquid nitrogen as the cryogenic medium. The results are shown in Figure~\ref{fig:VergleichElasto}b, together with the original measurement. Both measurements exhibit a high degree of agreement. The small deviations observed can likely be attributed to slight misalignment of the samples during mounting.

The fit parameters extracted from the different measurements are summarized in Table~\ref{tab:Fitparameter_BaFe2As2}. These results further confirm the good agreement between the different measurement approaches. The parameter $\lambda/a_0$ serves as a measure of the strength of nematic fluctuations. The largest deviation is observed between the measurement of sample 1 using the strain cell and the one obtained via the conventional method, but this deviation amounts to only 8\,\%. In contrast, the two measurements performed with the strain cell show excellent agreement, with a relative deviation of only 3.6\,\% in $\lambda/a_0$. The transition temperature $T_{\mathrm{nem}}$ also shows the largest discrepancy between the strain cell and the conventional measurement, with a relative deviation of just 2.5\,\%. Between the two strain cell measurements, the relative deviation in $T_{\mathrm{nem}}$ is only 0.5\,\%, further highlighting the reproducibility of the results. The fact that all extracted values for $T_{\mathrm{nem}}$ lie below the known transition temperature of $T_{\mathrm{nem,lit}} = 138$\,K is attributed to the coupling between the electronic nematic order and the crystal lattice~\cite{Kuo1}. The parameter $n_0$ reflects the intrinsic piezoresistive background and is not related to the electronic nematicity. Here too, only minor variations are observed across the different measurement approaches.


\begin{table}[]
\caption{Fit parameters from the Curie-Weiss fits to the elastoresistance measurements of BaFe$_2$As$_2$ performed using the strain cell and the conventional piezo-based method.}
\centering
\begin{tabular}{c|c|c|c}
measurement method & $\lambda /a_0$ (K) & $T_{\text{nem}}$ (K) & $n_0$  \\ \hline \hline
strain cell sample 1  &   $481\pm 3$  & $135.43 \pm  0.01$   & $3.46 \pm 0.04 $   \\ \hline 
strain cell sample 2  &   $499\pm 4$  & $134.73 \pm  0.04$   & $2.47 \pm 0.06 $   \\ \hline 
piezo actuator  &  $521 \pm 29 $   & $131.9 \pm 0.4 $   & $4.1 \pm 0.4 $   \\ 
\end{tabular}
\label{tab:Fitparameter_BaFe2As2}
\end{table}

\subsection{\label{sec:level7} Summary}
In this work, the construction of a cryo-compatible uniaxial strain cell consisting of three piezoelectric actuators was presented. The device was designed for various strain-dependent transport measurements. The maximum change in sample length reached -37\,µm under compression and +33\,µm under tension, corresponding to a strain of approximately $\pm 5.5\,\%$. Displacement was measured precisely using a high-resolution capacitive displacement sensor with high temporal resolution. Compared to conventional designs, the presented strain cell offers several advantages: the large-format piezoelectric actuators with a cross-sectional area of $7\,\mathrm{mm} \times 7\,\mathrm{mm}$ provide excellent mechanical stability and allow for large displacement ranges. The geometry of the strain cell offers an extended mounting area, enabling comfortable sample placement and flexible contact configuration. In addition, the high time resolution of the displacement sensor allows for dynamic strain tracking, which is not accessible with standard strain cell setups. \\
To enable measurements at low temperatures, a modular probe was developed into which the strain cell could be integrated. The probe was designed for cooling with liquid helium down to $4.2\,\mathrm{K}$. With an outer diameter of $49.5\,\mathrm{mm}$, the probe was compact enough to fit into cryostats with a clear inner opening of $50\,\mathrm{mm}$, thereby allowing, in principle, measurements under an external magnetic field.
For the operation of the strain cell in elastoresistivity measurements, an automated measurement program with a graphical user interface was developed, allowing simultaneous monitoring and control of the measurement process. The functionality of the entire system was demonstrated by elastoresistance measurements on BaFe$_2$As$_2$. The results were compared to those obtained using the conventional method, in which the sample is glued directly onto a single piezo actuator. Both approaches yielded consistent results, with significantly improved data quality when using the newly developed strain cell. Furthermore, the mechanical stability of the setup was confirmed by simultaneous operation of vacuum pumps, which did not have a significant effect on the data quality.\\
Beyond elastoresistivity, the strain cell could also be well suited for elastocaloric measurements, due to its high temporal resolution in strain detection and mechanical stability \cite{Ikeda2019}. In such experiments, the elastocaloric effect serves as a sensitive thermodynamic probe that enables the detection of strain-induced entropy changes and the investigation of phase transitions and associated fluctuations, including nematic correlations~\cite{Li2022,Ikeda2021}. Since elastocaloric signals are typically very weak and require precise, time-resolved strain tracking, the presented setup may offer advantages over conventional strain cells, which generally do not support dynamic strain detection. In previous experiments, this limitation was typically circumvented by indirectly estimating the strain amplitude from static calibration measurements---a procedure that could be associated with considerable uncertainties, especially when analyzing small temperature oscillations. In contrast, the strain cell developed in this work would allow for direct, time-resolved measurement of the actual strain during dynamic operation, potentially leading to a more accurate and reliable interpretation of elastocaloric data.

\begin{acknowledgments}
The author gratefully acknowledges Frederic Braun, Lukas Voss, and Florian Brockner for their assistance in setting up the experiment and for the many insightful discussions that greatly improved this work. Special thanks go to Jan Lino Kricke for his support during the mechanical loading analysis of the piezo actuators.
\end{acknowledgments}

\section*{Data Availability Statement}
The data that support the findings of this study are available from the corresponding author upon reasonable request.

\bibliography{Dehnung}

\begin{thebibliography}{28}%
\makeatletter
\providecommand \@ifxundefined [1]{%
 \@ifx{#1\undefined}
}%
\providecommand \@ifnum [1]{%
 \ifnum #1\expandafter \@firstoftwo
 \else \expandafter \@secondoftwo
 \fi
}%
\providecommand \@ifx [1]{%
 \ifx #1\expandafter \@firstoftwo
 \else \expandafter \@secondoftwo
 \fi
}%
\providecommand \natexlab [1]{#1}%
\providecommand \enquote  [1]{``#1''}%
\providecommand \bibnamefont  [1]{#1}%
\providecommand \bibfnamefont [1]{#1}%
\providecommand \citenamefont [1]{#1}%
\providecommand \href@noop [0]{\@secondoftwo}%
\providecommand \href [0]{\begingroup \@sanitize@url \@href}%
\providecommand \@href[1]{\@@startlink{#1}\@@href}%
\providecommand \@@href[1]{\endgroup#1\@@endlink}%
\providecommand \@sanitize@url [0]{\catcode `\\12\catcode `\$12\catcode `\&12\catcode `\#12\catcode `\^12\catcode `\_12\catcode `\%12\relax}%
\providecommand \@@startlink[1]{}%
\providecommand \@@endlink[0]{}%
\providecommand \url  [0]{\begingroup\@sanitize@url \@url }%
\providecommand \@url [1]{\endgroup\@href {#1}{\urlprefix }}%
\providecommand \urlprefix  [0]{URL }%
\providecommand \Eprint [0]{\href }%
\providecommand \doibase [0]{http://dx.doi.org/}%
\providecommand \selectlanguage [0]{\@gobble}%
\providecommand \bibinfo  [0]{\@secondoftwo}%
\providecommand \bibfield  [0]{\@secondoftwo}%
\providecommand \translation [1]{[#1]}%
\providecommand \BibitemOpen [0]{}%
\providecommand \bibitemStop [0]{}%
\providecommand \bibitemNoStop [0]{.\EOS\space}%
\providecommand \EOS [0]{\spacefactor3000\relax}%
\providecommand \BibitemShut  [1]{\csname bibitem#1\endcsname}%
\let\auto@bib@innerbib\@empty
\bibitem [{\citenamefont {Shirakawa}\ \emph {et~al.}(1997)\citenamefont {Shirakawa}, \citenamefont {Murata}, \citenamefont {Nishizaki}, \citenamefont {Maeno},\ and\ \citenamefont {Fujita}}]{Dehnungszelle11}%
  \BibitemOpen
  \bibfield  {author} {\bibinfo {author} {\bibfnamefont {N.}~\bibnamefont {Shirakawa}}, \bibinfo {author} {\bibfnamefont {K.}~\bibnamefont {Murata}}, \bibinfo {author} {\bibfnamefont {S.}~\bibnamefont {Nishizaki}}, \bibinfo {author} {\bibfnamefont {Y.}~\bibnamefont {Maeno}}, \ and\ \bibinfo {author} {\bibfnamefont {T.}~\bibnamefont {Fujita}},\ }\bibfield  {title} {\enquote {\bibinfo {title} {Pressure dependence of superconducting critical temperature of {Sr}$_2${RuO}$_4$},}\ }\href {\doibase 10.1103/PhysRevB.56.7890} {\bibfield  {journal} {\bibinfo  {journal} {Phys. Rev. B}\ }\textbf {\bibinfo {volume} {56}},\ \bibinfo {pages} {7890--7893} (\bibinfo {year} {1997})}\BibitemShut {NoStop}%
\bibitem [{\citenamefont {Hicks}\ \emph {et~al.}(2014{\natexlab{a}})\citenamefont {Hicks}, \citenamefont {Brodsky}, \citenamefont {Yelland}, \citenamefont {Gibbs}, \citenamefont {Bruin}, \citenamefont {Barber}, \citenamefont {Edkins}, \citenamefont {Nishimura}, \citenamefont {Yonezawa}, \citenamefont {Maeno},\ and\ \citenamefont {Mackenzie}}]{Dehnungszelle21}%
  \BibitemOpen
  \bibfield  {author} {\bibinfo {author} {\bibfnamefont {C.~W.}\ \bibnamefont {Hicks}}, \bibinfo {author} {\bibfnamefont {D.~O.}\ \bibnamefont {Brodsky}}, \bibinfo {author} {\bibfnamefont {E.~A.}\ \bibnamefont {Yelland}}, \bibinfo {author} {\bibfnamefont {A.~S.}\ \bibnamefont {Gibbs}}, \bibinfo {author} {\bibfnamefont {J.~A.~N.}\ \bibnamefont {Bruin}}, \bibinfo {author} {\bibfnamefont {M.~E.}\ \bibnamefont {Barber}}, \bibinfo {author} {\bibfnamefont {S.~D.}\ \bibnamefont {Edkins}}, \bibinfo {author} {\bibfnamefont {K.}~\bibnamefont {Nishimura}}, \bibinfo {author} {\bibfnamefont {S.}~\bibnamefont {Yonezawa}}, \bibinfo {author} {\bibfnamefont {Y.}~\bibnamefont {Maeno}}, \ and\ \bibinfo {author} {\bibfnamefont {A.~P.}\ \bibnamefont {Mackenzie}},\ }\bibfield  {title} {\enquote {\bibinfo {title} {Strong increase of $t_c$ of {Sr}$_2${RuO}$_4$ under both tensile and compressive strain},}\ }\href {\doibase 10.1126/science.1248292} {\bibfield  {journal} {\bibinfo  {journal} {Science}\ }\textbf {\bibinfo {volume}
  {344}},\ \bibinfo {pages} {283--285} (\bibinfo {year} {2014}{\natexlab{a}})},\ \Eprint {http://arxiv.org/abs/https://www.science.org/doi/pdf/10.1126/science.1248292} {https://www.science.org/doi/pdf/10.1126/science.1248292} \BibitemShut {NoStop}%
\bibitem [{\citenamefont {Taniguchi}\ \emph {et~al.}(2015)\citenamefont {Taniguchi}, \citenamefont {Nishimura}, \citenamefont {Goh}, \citenamefont {Yonezawa},\ and\ \citenamefont {Maeno}}]{Dehnungszelle3}%
  \BibitemOpen
  \bibfield  {author} {\bibinfo {author} {\bibfnamefont {H.}~\bibnamefont {Taniguchi}}, \bibinfo {author} {\bibfnamefont {K.}~\bibnamefont {Nishimura}}, \bibinfo {author} {\bibfnamefont {S.~K.}\ \bibnamefont {Goh}}, \bibinfo {author} {\bibfnamefont {S.}~\bibnamefont {Yonezawa}}, \ and\ \bibinfo {author} {\bibfnamefont {Y.}~\bibnamefont {Maeno}},\ }\bibfield  {title} {\enquote {\bibinfo {title} {Higher-$t_c$ superconducting phase in {Sr}$_2${RuO}$_4$ induced by in-plane uniaxial pressure},}\ }\href {\doibase 10.7566/JPSJ.84.014707} {\bibfield  {journal} {\bibinfo  {journal} {Journal of the Physical Society of Japan}\ }\textbf {\bibinfo {volume} {84}},\ \bibinfo {pages} {014707} (\bibinfo {year} {2015})},\ \Eprint {http://arxiv.org/abs/https://doi.org/10.7566/JPSJ.84.014707} {https://doi.org/10.7566/JPSJ.84.014707} \BibitemShut {NoStop}%
\bibitem [{\citenamefont {Sadewasser}\ \emph {et~al.}(2000)\citenamefont {Sadewasser}, \citenamefont {Schilling}, \citenamefont {Paulikas},\ and\ \citenamefont {Veal}}]{Dehnungszelle4}%
  \BibitemOpen
  \bibfield  {author} {\bibinfo {author} {\bibfnamefont {S.}~\bibnamefont {Sadewasser}}, \bibinfo {author} {\bibfnamefont {J.~S.}\ \bibnamefont {Schilling}}, \bibinfo {author} {\bibfnamefont {A.~P.}\ \bibnamefont {Paulikas}}, \ and\ \bibinfo {author} {\bibfnamefont {B.~W.}\ \bibnamefont {Veal}},\ }\bibfield  {title} {\enquote {\bibinfo {title} {Pressure dependence of $t_c$ to 17 {GPa} with and without relaxation effects in superconducting {YBa}$_2${Cu}$_3${O}$_x$},}\ }\href {\doibase 10.1103/PhysRevB.61.741} {\bibfield  {journal} {\bibinfo  {journal} {Phys. Rev. B}\ }\textbf {\bibinfo {volume} {61}},\ \bibinfo {pages} {741--749} (\bibinfo {year} {2000})}\BibitemShut {NoStop}%
\bibitem [{\citenamefont {Kim}\ \emph {et~al.}(2018)\citenamefont {Kim}, \citenamefont {Souliou}, \citenamefont {Barber}, \citenamefont {Lefrançois}, \citenamefont {Minola}, \citenamefont {Tortora}, \citenamefont {Heid}, \citenamefont {Nandi}, \citenamefont {Borzi}, \citenamefont {Garbarino}, \citenamefont {Bosak}, \citenamefont {Porras}, \citenamefont {Loew}, \citenamefont {König}, \citenamefont {Moll}, \citenamefont {Mackenzie}, \citenamefont {Keimer}, \citenamefont {Hicks},\ and\ \citenamefont {Tacon}}]{Dehnungszelle5}%
  \BibitemOpen
  \bibfield  {author} {\bibinfo {author} {\bibfnamefont {H.-H.}\ \bibnamefont {Kim}}, \bibinfo {author} {\bibfnamefont {S.~M.}\ \bibnamefont {Souliou}}, \bibinfo {author} {\bibfnamefont {M.~E.}\ \bibnamefont {Barber}}, \bibinfo {author} {\bibfnamefont {E.}~\bibnamefont {Lefrançois}}, \bibinfo {author} {\bibfnamefont {M.}~\bibnamefont {Minola}}, \bibinfo {author} {\bibfnamefont {M.}~\bibnamefont {Tortora}}, \bibinfo {author} {\bibfnamefont {R.}~\bibnamefont {Heid}}, \bibinfo {author} {\bibfnamefont {N.}~\bibnamefont {Nandi}}, \bibinfo {author} {\bibfnamefont {R.~A.}\ \bibnamefont {Borzi}}, \bibinfo {author} {\bibfnamefont {G.}~\bibnamefont {Garbarino}}, \bibinfo {author} {\bibfnamefont {A.}~\bibnamefont {Bosak}}, \bibinfo {author} {\bibfnamefont {J.}~\bibnamefont {Porras}}, \bibinfo {author} {\bibfnamefont {T.}~\bibnamefont {Loew}}, \bibinfo {author} {\bibfnamefont {M.}~\bibnamefont {König}}, \bibinfo {author} {\bibfnamefont {P.~J.~W.}\ \bibnamefont {Moll}}, \bibinfo {author} {\bibfnamefont {A.~P.}\
  \bibnamefont {Mackenzie}}, \bibinfo {author} {\bibfnamefont {B.}~\bibnamefont {Keimer}}, \bibinfo {author} {\bibfnamefont {C.~W.}\ \bibnamefont {Hicks}}, \ and\ \bibinfo {author} {\bibfnamefont {M.~L.}\ \bibnamefont {Tacon}},\ }\bibfield  {title} {\enquote {\bibinfo {title} {Uniaxial pressure control of competing orders in a high-temperature superconductor},}\ }\href {\doibase 10.1126/science.aat4708} {\bibfield  {journal} {\bibinfo  {journal} {Science}\ }\textbf {\bibinfo {volume} {362}},\ \bibinfo {pages} {1040--1044} (\bibinfo {year} {2018})},\ \Eprint {http://arxiv.org/abs/https://www.science.org/doi/pdf/10.1126/science.aat4708} {https://www.science.org/doi/pdf/10.1126/science.aat4708} \BibitemShut {NoStop}%
\bibitem [{\citenamefont {Tanatar}\ \emph {et~al.}(2010)\citenamefont {Tanatar}, \citenamefont {Blomberg}, \citenamefont {Kreyssig}, \citenamefont {Kim}, \citenamefont {Ni}, \citenamefont {Thaler}, \citenamefont {Bud'ko}, \citenamefont {Canfield}, \citenamefont {Goldman}, \citenamefont {Mazin},\ and\ \citenamefont {Prozorov}}]{MechanischStrain1}%
  \BibitemOpen
  \bibfield  {author} {\bibinfo {author} {\bibfnamefont {M.~A.}\ \bibnamefont {Tanatar}}, \bibinfo {author} {\bibfnamefont {E.~C.}\ \bibnamefont {Blomberg}}, \bibinfo {author} {\bibfnamefont {A.}~\bibnamefont {Kreyssig}}, \bibinfo {author} {\bibfnamefont {M.~G.}\ \bibnamefont {Kim}}, \bibinfo {author} {\bibfnamefont {N.}~\bibnamefont {Ni}}, \bibinfo {author} {\bibfnamefont {A.}~\bibnamefont {Thaler}}, \bibinfo {author} {\bibfnamefont {S.~L.}\ \bibnamefont {Bud'ko}}, \bibinfo {author} {\bibfnamefont {P.~C.}\ \bibnamefont {Canfield}}, \bibinfo {author} {\bibfnamefont {A.~I.}\ \bibnamefont {Goldman}}, \bibinfo {author} {\bibfnamefont {I.~I.}\ \bibnamefont {Mazin}}, \ and\ \bibinfo {author} {\bibfnamefont {R.}~\bibnamefont {Prozorov}},\ }\bibfield  {title} {\enquote {\bibinfo {title} {Uniaxial-strain mechanical detwinning of {CaFe}$_2${As}$_2$ and {BaFe}$_2${As}$_2$ crystals: Optical and transport study},}\ }\href {\doibase 10.1103/PhysRevB.81.184508} {\bibfield  {journal} {\bibinfo  {journal} {Phys. Rev. B}\
  }\textbf {\bibinfo {volume} {81}},\ \bibinfo {pages} {184508} (\bibinfo {year} {2010})}\BibitemShut {NoStop}%
\bibitem [{\citenamefont {Dhital}\ \emph {et~al.}(2012)\citenamefont {Dhital}, \citenamefont {Yamani}, \citenamefont {Tian}, \citenamefont {Zeretsky}, \citenamefont {Sefat}, \citenamefont {Wang}, \citenamefont {Birgeneau},\ and\ \citenamefont {Wilson}}]{MechanischStrain2}%
  \BibitemOpen
  \bibfield  {author} {\bibinfo {author} {\bibfnamefont {C.}~\bibnamefont {Dhital}}, \bibinfo {author} {\bibfnamefont {Z.}~\bibnamefont {Yamani}}, \bibinfo {author} {\bibfnamefont {W.}~\bibnamefont {Tian}}, \bibinfo {author} {\bibfnamefont {J.}~\bibnamefont {Zeretsky}}, \bibinfo {author} {\bibfnamefont {A.~S.}\ \bibnamefont {Sefat}}, \bibinfo {author} {\bibfnamefont {Z.}~\bibnamefont {Wang}}, \bibinfo {author} {\bibfnamefont {R.~J.}\ \bibnamefont {Birgeneau}}, \ and\ \bibinfo {author} {\bibfnamefont {S.~D.}\ \bibnamefont {Wilson}},\ }\bibfield  {title} {\enquote {\bibinfo {title} {Effect of uniaxial strain on the structural and magnetic phase transitions in {BaFe}$_2${As}$_2$},}\ }\href {\doibase 10.1103/PhysRevLett.108.087001} {\bibfield  {journal} {\bibinfo  {journal} {Phys. Rev. Lett.}\ }\textbf {\bibinfo {volume} {108}},\ \bibinfo {pages} {087001} (\bibinfo {year} {2012})}\BibitemShut {NoStop}%
\bibitem [{\citenamefont {Ren}\ \emph {et~al.}(2015)\citenamefont {Ren}, \citenamefont {Duan}, \citenamefont {Hu}, \citenamefont {Li}, \citenamefont {Zhang}, \citenamefont {Luo}, \citenamefont {Dai},\ and\ \citenamefont {Li}}]{MechanischStrain3}%
  \BibitemOpen
  \bibfield  {author} {\bibinfo {author} {\bibfnamefont {X.}~\bibnamefont {Ren}}, \bibinfo {author} {\bibfnamefont {L.}~\bibnamefont {Duan}}, \bibinfo {author} {\bibfnamefont {Y.}~\bibnamefont {Hu}}, \bibinfo {author} {\bibfnamefont {J.}~\bibnamefont {Li}}, \bibinfo {author} {\bibfnamefont {R.}~\bibnamefont {Zhang}}, \bibinfo {author} {\bibfnamefont {H.}~\bibnamefont {Luo}}, \bibinfo {author} {\bibfnamefont {P.}~\bibnamefont {Dai}}, \ and\ \bibinfo {author} {\bibfnamefont {Y.}~\bibnamefont {Li}},\ }\bibfield  {title} {\enquote {\bibinfo {title} {Nematic crossover in {BaFe}$_2${As}$_2$ under uniaxial stress},}\ }\href {\doibase 10.1103/PhysRevLett.115.197002} {\bibfield  {journal} {\bibinfo  {journal} {Phys. Rev. Lett.}\ }\textbf {\bibinfo {volume} {115}},\ \bibinfo {pages} {197002} (\bibinfo {year} {2015})}\BibitemShut {NoStop}%
\bibitem [{\citenamefont {Chu}\ \emph {et~al.}(2010{\natexlab{a}})\citenamefont {Chu}, \citenamefont {Analytis}, \citenamefont {Greve}, \citenamefont {McMahon}, \citenamefont {Islam}, \citenamefont {Yamamoto},\ and\ \citenamefont {Fisher}}]{Chu2010}%
  \BibitemOpen
  \bibfield  {author} {\bibinfo {author} {\bibfnamefont {J.-H.}\ \bibnamefont {Chu}}, \bibinfo {author} {\bibfnamefont {J.~G.}\ \bibnamefont {Analytis}}, \bibinfo {author} {\bibfnamefont {K.~D.}\ \bibnamefont {Greve}}, \bibinfo {author} {\bibfnamefont {P.~L.}\ \bibnamefont {McMahon}}, \bibinfo {author} {\bibfnamefont {Z.}~\bibnamefont {Islam}}, \bibinfo {author} {\bibfnamefont {Y.}~\bibnamefont {Yamamoto}}, \ and\ \bibinfo {author} {\bibfnamefont {I.~R.}\ \bibnamefont {Fisher}},\ }\bibfield  {title} {\enquote {\bibinfo {title} {In-plane resistivity anisotropy in an underdoped iron arsenide superconductor},}\ }\href {\doibase 10.1126/science.1190482} {\bibfield  {journal} {\bibinfo  {journal} {Science}\ }\textbf {\bibinfo {volume} {329}},\ \bibinfo {pages} {824--826} (\bibinfo {year} {2010}{\natexlab{a}})},\ \Eprint {http://arxiv.org/abs/https://www.science.org/doi/pdf/10.1126/science.1190482} {https://www.science.org/doi/pdf/10.1126/science.1190482} \BibitemShut {NoStop}%
\bibitem [{\citenamefont {Degiorgi}(2022)}]{BlasBalck}%
  \BibitemOpen
  \bibfield  {author} {\bibinfo {author} {\bibfnamefont {L.}~\bibnamefont {Degiorgi}},\ }\bibfield  {title} {\enquote {\bibinfo {title} {Optical fingerprints of nematicity in iron-based superconductors},}\ }\href {\doibase 10.3389/fphy.2022.866664} {\bibfield  {journal} {\bibinfo  {journal} {Frontiers in Physics}\ }\textbf {\bibinfo {volume} {10}} (\bibinfo {year} {2022}),\ 10.3389/fphy.2022.866664}\BibitemShut {NoStop}%
\bibitem [{\citenamefont {Conley}\ \emph {et~al.}(2013)\citenamefont {Conley}, \citenamefont {Wang}, \citenamefont {Ziegler}, \citenamefont {Haglund~Jr.}, \citenamefont {Pantelides},\ and\ \citenamefont {Bolotin}}]{Biegen11}%
  \BibitemOpen
  \bibfield  {author} {\bibinfo {author} {\bibfnamefont {H.~J.}\ \bibnamefont {Conley}}, \bibinfo {author} {\bibfnamefont {B.}~\bibnamefont {Wang}}, \bibinfo {author} {\bibfnamefont {J.~I.}\ \bibnamefont {Ziegler}}, \bibinfo {author} {\bibfnamefont {R.~F.}\ \bibnamefont {Haglund~Jr.}}, \bibinfo {author} {\bibfnamefont {S.~T.}\ \bibnamefont {Pantelides}}, \ and\ \bibinfo {author} {\bibfnamefont {K.~I.}\ \bibnamefont {Bolotin}},\ }\bibfield  {title} {\enquote {\bibinfo {title} {Bandgap engineering of strained monolayer and bilayer {MoS}$_2$},}\ }\href {\doibase 10.1021/nl4014748} {\bibfield  {journal} {\bibinfo  {journal} {Nano Letters}\ }\textbf {\bibinfo {volume} {13}},\ \bibinfo {pages} {3626--3630} (\bibinfo {year} {2013})}\BibitemShut {NoStop}%
\bibitem [{\citenamefont {Mohiuddin}\ \emph {et~al.}(2009)\citenamefont {Mohiuddin}, \citenamefont {Lombardo}, \citenamefont {Nair}, \citenamefont {Bonetti}, \citenamefont {Savini}, \citenamefont {Jalil}, \citenamefont {Bonini}, \citenamefont {Basko}, \citenamefont {Galiotis}, \citenamefont {Marzari}, \citenamefont {Novoselov}, \citenamefont {Geim},\ and\ \citenamefont {Ferrari}}]{Biegen21}%
  \BibitemOpen
  \bibfield  {author} {\bibinfo {author} {\bibfnamefont {T.~M.~G.}\ \bibnamefont {Mohiuddin}}, \bibinfo {author} {\bibfnamefont {A.}~\bibnamefont {Lombardo}}, \bibinfo {author} {\bibfnamefont {R.~R.}\ \bibnamefont {Nair}}, \bibinfo {author} {\bibfnamefont {A.}~\bibnamefont {Bonetti}}, \bibinfo {author} {\bibfnamefont {G.}~\bibnamefont {Savini}}, \bibinfo {author} {\bibfnamefont {R.}~\bibnamefont {Jalil}}, \bibinfo {author} {\bibfnamefont {N.}~\bibnamefont {Bonini}}, \bibinfo {author} {\bibfnamefont {D.~M.}\ \bibnamefont {Basko}}, \bibinfo {author} {\bibfnamefont {C.}~\bibnamefont {Galiotis}}, \bibinfo {author} {\bibfnamefont {N.}~\bibnamefont {Marzari}}, \bibinfo {author} {\bibfnamefont {K.~S.}\ \bibnamefont {Novoselov}}, \bibinfo {author} {\bibfnamefont {A.~K.}\ \bibnamefont {Geim}}, \ and\ \bibinfo {author} {\bibfnamefont {A.~C.}\ \bibnamefont {Ferrari}},\ }\bibfield  {title} {\enquote {\bibinfo {title} {Uniaxial strain in graphene by raman spectroscopy: {$G$} peak splitting, {G}r\"uneisen parameters, and
  sample orientation},}\ }\href {\doibase 10.1103/PhysRevB.79.205433} {\bibfield  {journal} {\bibinfo  {journal} {Phys. Rev. B}\ }\textbf {\bibinfo {volume} {79}},\ \bibinfo {pages} {205433} (\bibinfo {year} {2009})}\BibitemShut {NoStop}%
\bibitem [{\citenamefont {Hicks}\ \emph {et~al.}(2014{\natexlab{b}})\citenamefont {Hicks}, \citenamefont {Barber}, \citenamefont {Edkins}, \citenamefont {Brodsky},\ and\ \citenamefont {Mackenzie}}]{Dehnungszelle1}%
  \BibitemOpen
  \bibfield  {author} {\bibinfo {author} {\bibfnamefont {C.~W.}\ \bibnamefont {Hicks}}, \bibinfo {author} {\bibfnamefont {M.~E.}\ \bibnamefont {Barber}}, \bibinfo {author} {\bibfnamefont {S.~D.}\ \bibnamefont {Edkins}}, \bibinfo {author} {\bibfnamefont {D.~O.}\ \bibnamefont {Brodsky}}, \ and\ \bibinfo {author} {\bibfnamefont {A.~P.}\ \bibnamefont {Mackenzie}},\ }\bibfield  {title} {\enquote {\bibinfo {title} {{Piezoelectric-based apparatus for strain tuning}},}\ }\href {\doibase 10.1063/1.4881611} {\bibfield  {journal} {\bibinfo  {journal} {Review of Scientific Instruments}\ }\textbf {\bibinfo {volume} {85}},\ \bibinfo {pages} {065003} (\bibinfo {year} {2014}{\natexlab{b}})},\ \Eprint {http://arxiv.org/abs/https://pubs.aip.org/aip/rsi/article-pdf/doi/10.1063/1.4881611/16010151/065003\_1\_online.pdf} {https://pubs.aip.org/aip/rsi/article-pdf/doi/10.1063/1.4881611/16010151/065003\_1\_online.pdf} \BibitemShut {NoStop}%
\bibitem [{\citenamefont {Martienssen}\ and\ \citenamefont {Warlimont}(2005)}]{martienssen2005springer}%
  \BibitemOpen
  \bibfield  {author} {\bibinfo {author} {\bibfnamefont {W.}~\bibnamefont {Martienssen}}\ and\ \bibinfo {author} {\bibfnamefont {H.}~\bibnamefont {Warlimont}},\ }\href {https://books.google.de/books?id=ELnUngEACAAJ} {\emph {\bibinfo {title} {Springer Handbook of Condensed Matter and Materials Data: CD-ROM}}},\ Springer Handbook of Condensed Matter and Materials Data\ (\bibinfo  {publisher} {Springer},\ \bibinfo {year} {2005})\BibitemShut {NoStop}%
\bibitem [{\citenamefont {{HEMPEL metals and more}}(2024)}]{Titan}%
  \BibitemOpen
  \bibfield  {author} {\bibinfo {author} {\bibnamefont {{HEMPEL metals and more}}},\ }\href@noop {} {\enquote {\bibinfo {title} {{Titan grade 2 - 3.7025, UNS R50400, Werkstoff, Material, Metall}},}\ } (\bibinfo {year} {2024}),\ \bibinfo {note} {\url{https://www.hempel-metals.de/de/werkstoffe/titanlegierungen/ti-gr-2-37035/} [Accessed: (20.06.2024)]}\BibitemShut {NoStop}%
\bibitem [{\citenamefont {{MicroEpsilon}}()}]{Abstandssensor}%
  \BibitemOpen
  \bibfield  {author} {\bibinfo {author} {\bibnamefont {{MicroEpsilon}}},\ }\href@noop {} {\emph {\bibinfo {title} {capaNCDT, Kapazitive Sensoren für Weg, Abstand und Position}}}\BibitemShut {NoStop}%
\bibitem [{\citenamefont {Hong}\ \emph {et~al.}(2022)\citenamefont {Hong}, \citenamefont {Sykora}, \citenamefont {Caglieris}, \citenamefont {Behnami}, \citenamefont {Morozov}, \citenamefont {Aswartham}, \citenamefont {Grinenko}, \citenamefont {Kihou}, \citenamefont {Lee}, \citenamefont {Büchner},\ and\ \citenamefont {Hess}}]{Xiaochen2}%
  \BibitemOpen
  \bibfield  {author} {\bibinfo {author} {\bibfnamefont {X.}~\bibnamefont {Hong}}, \bibinfo {author} {\bibfnamefont {S.}~\bibnamefont {Sykora}}, \bibinfo {author} {\bibfnamefont {F.}~\bibnamefont {Caglieris}}, \bibinfo {author} {\bibfnamefont {M.}~\bibnamefont {Behnami}}, \bibinfo {author} {\bibfnamefont {I.}~\bibnamefont {Morozov}}, \bibinfo {author} {\bibfnamefont {S.}~\bibnamefont {Aswartham}}, \bibinfo {author} {\bibfnamefont {V.}~\bibnamefont {Grinenko}}, \bibinfo {author} {\bibfnamefont {K.}~\bibnamefont {Kihou}}, \bibinfo {author} {\bibfnamefont {C.-H.}\ \bibnamefont {Lee}}, \bibinfo {author} {\bibfnamefont {B.}~\bibnamefont {Büchner}}, \ and\ \bibinfo {author} {\bibfnamefont {C.}~\bibnamefont {Hess}},\ }\bibfield  {title} {\enquote {\bibinfo {title} {Elastoresistivity of heavily hole-doped 122 iron pnictide superconductors},}\ }\href {\doibase 10.3389/fphy.2022.853717} {\bibfield  {journal} {\bibinfo  {journal} {Frontiers in Physics}\ }\textbf {\bibinfo {volume} {10}} (\bibinfo {year} {2022}),\
  10.3389/fphy.2022.853717}\BibitemShut {NoStop}%
\bibitem [{\citenamefont {Hong}\ \emph {et~al.}(2020)\citenamefont {Hong}, \citenamefont {Caglieris}, \citenamefont {Kappenberger}, \citenamefont {Wurmehl}, \citenamefont {Aswartham}, \citenamefont {Scaravaggi}, \citenamefont {Lepucki}, \citenamefont {Wolter}, \citenamefont {Grafe}, \citenamefont {B\"uchner},\ and\ \citenamefont {Hess}}]{Xiaochen1}%
  \BibitemOpen
  \bibfield  {author} {\bibinfo {author} {\bibfnamefont {X.}~\bibnamefont {Hong}}, \bibinfo {author} {\bibfnamefont {F.}~\bibnamefont {Caglieris}}, \bibinfo {author} {\bibfnamefont {R.}~\bibnamefont {Kappenberger}}, \bibinfo {author} {\bibfnamefont {S.}~\bibnamefont {Wurmehl}}, \bibinfo {author} {\bibfnamefont {S.}~\bibnamefont {Aswartham}}, \bibinfo {author} {\bibfnamefont {F.}~\bibnamefont {Scaravaggi}}, \bibinfo {author} {\bibfnamefont {P.}~\bibnamefont {Lepucki}}, \bibinfo {author} {\bibfnamefont {A.~U.~B.}\ \bibnamefont {Wolter}}, \bibinfo {author} {\bibfnamefont {H.-J.}\ \bibnamefont {Grafe}}, \bibinfo {author} {\bibfnamefont {B.}~\bibnamefont {B\"uchner}}, \ and\ \bibinfo {author} {\bibfnamefont {C.}~\bibnamefont {Hess}},\ }\bibfield  {title} {\enquote {\bibinfo {title} {Evolution of the nematic susceptibility in {LaFe}$_{1-x}${Co}$_x${AsO}},}\ }\href {\doibase 10.1103/PhysRevLett.125.067001} {\bibfield  {journal} {\bibinfo  {journal} {Phys. Rev. Lett.}\ }\textbf {\bibinfo {volume} {125}},\ \bibinfo
  {pages} {067001} (\bibinfo {year} {2020})}\BibitemShut {NoStop}%
\bibitem [{\citenamefont {Fernandes}, \citenamefont {Chubukov},\ and\ \citenamefont {Schmalian}(2014)}]{Fernandes2014}%
  \BibitemOpen
  \bibfield  {author} {\bibinfo {author} {\bibfnamefont {R.~M.}\ \bibnamefont {Fernandes}}, \bibinfo {author} {\bibfnamefont {A.~V.}\ \bibnamefont {Chubukov}}, \ and\ \bibinfo {author} {\bibfnamefont {J.}~\bibnamefont {Schmalian}},\ }\bibfield  {title} {\enquote {\bibinfo {title} {What drives nematic order in iron-based superconductors?}}\ }\href {\doibase 10.1038/nphys2877} {\bibfield  {journal} {\bibinfo  {journal} {Nature Physics}\ }\textbf {\bibinfo {volume} {10}},\ \bibinfo {pages} {1745--2481} (\bibinfo {year} {2014})}\BibitemShut {NoStop}%
\bibitem [{\citenamefont {Fernandes}\ \emph {et~al.}(2013)\citenamefont {Fernandes}, \citenamefont {B\"ohmer}, \citenamefont {Meingast},\ and\ \citenamefont {Schmalian}}]{Fernandes2013}%
  \BibitemOpen
  \bibfield  {author} {\bibinfo {author} {\bibfnamefont {R.~M.}\ \bibnamefont {Fernandes}}, \bibinfo {author} {\bibfnamefont {A.~E.}\ \bibnamefont {B\"ohmer}}, \bibinfo {author} {\bibfnamefont {C.}~\bibnamefont {Meingast}}, \ and\ \bibinfo {author} {\bibfnamefont {J.}~\bibnamefont {Schmalian}},\ }\bibfield  {title} {\enquote {\bibinfo {title} {Scaling between magnetic and lattice fluctuations in iron pnictide superconductors},}\ }\href {\doibase 10.1103/PhysRevLett.111.137001} {\bibfield  {journal} {\bibinfo  {journal} {Phys. Rev. Lett.}\ }\textbf {\bibinfo {volume} {111}},\ \bibinfo {pages} {137001} (\bibinfo {year} {2013})}\BibitemShut {NoStop}%
\bibitem [{\citenamefont {Canfield}\ and\ \citenamefont {Bud'ko}(2010)}]{test}%
  \BibitemOpen
  \bibfield  {author} {\bibinfo {author} {\bibfnamefont {P.~C.}\ \bibnamefont {Canfield}}\ and\ \bibinfo {author} {\bibfnamefont {S.~L.}\ \bibnamefont {Bud'ko}},\ }\bibfield  {title} {\enquote {\bibinfo {title} {{FeAs}-based superconductivity: A case study of the effects of transition metal doping on bafe$_2$as$_2$},}\ }\href {\doibase 10.1146/annurev-conmatphys-070909-104041} {\bibfield  {journal} {\bibinfo  {journal} {Annual Review of Condensed Matter Physics}\ }\textbf {\bibinfo {volume} {1}},\ \bibinfo {pages} {27--50} (\bibinfo {year} {2010})}\BibitemShut {NoStop}%
\bibitem [{\citenamefont {Chu}\ \emph {et~al.}(2012)\citenamefont {Chu}, \citenamefont {Kuo}, \citenamefont {Analytis},\ and\ \citenamefont {Fisher}}]{Chu}%
  \BibitemOpen
  \bibfield  {author} {\bibinfo {author} {\bibfnamefont {J.-H.}\ \bibnamefont {Chu}}, \bibinfo {author} {\bibfnamefont {H.-H.}\ \bibnamefont {Kuo}}, \bibinfo {author} {\bibfnamefont {J.~G.}\ \bibnamefont {Analytis}}, \ and\ \bibinfo {author} {\bibfnamefont {I.~R.}\ \bibnamefont {Fisher}},\ }\bibfield  {title} {\enquote {\bibinfo {title} {Divergent nematic susceptibility in an iron arsenide superconductor},}\ }\href {\doibase 10.1126/science.1221713} {\bibfield  {journal} {\bibinfo  {journal} {Science}\ }\textbf {\bibinfo {volume} {337}},\ \bibinfo {pages} {710--712} (\bibinfo {year} {2012})},\ \Eprint {http://arxiv.org/abs/https://www.science.org/doi/pdf/10.1126/science.1221713} {https://www.science.org/doi/pdf/10.1126/science.1221713} \BibitemShut {NoStop}%
\bibitem [{\citenamefont {Chu}\ \emph {et~al.}(2010{\natexlab{b}})\citenamefont {Chu}, \citenamefont {Analytis}, \citenamefont {Greve}, \citenamefont {McMahon}, \citenamefont {Islam}, \citenamefont {Yamamoto},\ and\ \citenamefont {Fisher}}]{Chualt}%
  \BibitemOpen
  \bibfield  {author} {\bibinfo {author} {\bibfnamefont {J.-H.}\ \bibnamefont {Chu}}, \bibinfo {author} {\bibfnamefont {J.~G.}\ \bibnamefont {Analytis}}, \bibinfo {author} {\bibfnamefont {K.~D.}\ \bibnamefont {Greve}}, \bibinfo {author} {\bibfnamefont {P.~L.}\ \bibnamefont {McMahon}}, \bibinfo {author} {\bibfnamefont {Z.}~\bibnamefont {Islam}}, \bibinfo {author} {\bibfnamefont {Y.}~\bibnamefont {Yamamoto}}, \ and\ \bibinfo {author} {\bibfnamefont {I.~R.}\ \bibnamefont {Fisher}},\ }\bibfield  {title} {\enquote {\bibinfo {title} {In-plane resistivity anisotropy in an underdoped iron arsenide superconductor},}\ }\href {\doibase 10.1126/science.1190482} {\bibfield  {journal} {\bibinfo  {journal} {Science}\ }\textbf {\bibinfo {volume} {329}},\ \bibinfo {pages} {824--826} (\bibinfo {year} {2010}{\natexlab{b}})},\ \Eprint {http://arxiv.org/abs/https://www.science.org/doi/pdf/10.1126/science.1190482} {https://www.science.org/doi/pdf/10.1126/science.1190482} \BibitemShut {NoStop}%
\bibitem [{\citenamefont {Kuo}\ \emph {et~al.}(2013)\citenamefont {Kuo}, \citenamefont {Shapiro}, \citenamefont {Riggs},\ and\ \citenamefont {Fisher}}]{Kuo1}%
  \BibitemOpen
  \bibfield  {author} {\bibinfo {author} {\bibfnamefont {H.-H.}\ \bibnamefont {Kuo}}, \bibinfo {author} {\bibfnamefont {M.~C.}\ \bibnamefont {Shapiro}}, \bibinfo {author} {\bibfnamefont {S.~C.}\ \bibnamefont {Riggs}}, \ and\ \bibinfo {author} {\bibfnamefont {I.~R.}\ \bibnamefont {Fisher}},\ }\bibfield  {title} {\enquote {\bibinfo {title} {Measurement of the elastoresistivity coefficients of the underdoped iron arsenide {Ba}({Fe}$_{0.975}${Co}$_{0.025}$)$_2${As}$_2$},}\ }\href {\doibase 10.1103/PhysRevB.88.085113} {\bibfield  {journal} {\bibinfo  {journal} {Phys. Rev. B}\ }\textbf {\bibinfo {volume} {88}},\ \bibinfo {pages} {085113} (\bibinfo {year} {2013})}\BibitemShut {NoStop}%
\bibitem [{\citenamefont {Kuo}\ \emph {et~al.}(2016)\citenamefont {Kuo}, \citenamefont {Chu}, \citenamefont {Palmstrom}, \citenamefont {Kivelson},\ and\ \citenamefont {Fisher}}]{Kuo2}%
  \BibitemOpen
  \bibfield  {author} {\bibinfo {author} {\bibfnamefont {H.-H.}\ \bibnamefont {Kuo}}, \bibinfo {author} {\bibfnamefont {J.-H.}\ \bibnamefont {Chu}}, \bibinfo {author} {\bibfnamefont {J.~C.}\ \bibnamefont {Palmstrom}}, \bibinfo {author} {\bibfnamefont {S.~A.}\ \bibnamefont {Kivelson}}, \ and\ \bibinfo {author} {\bibfnamefont {I.~R.}\ \bibnamefont {Fisher}},\ }\bibfield  {title} {\enquote {\bibinfo {title} {Ubiquitous signatures of nematic quantum criticality in optimally doped {Fe}-based superconductors},}\ }\href {\doibase 10.1126/science.aab0103} {\bibfield  {journal} {\bibinfo  {journal} {Science}\ }\textbf {\bibinfo {volume} {352}},\ \bibinfo {pages} {958--962} (\bibinfo {year} {2016})},\ \Eprint {http://arxiv.org/abs/https://www.science.org/doi/pdf/10.1126/science.aab0103} {https://www.science.org/doi/pdf/10.1126/science.aab0103} \BibitemShut {NoStop}%
\bibitem [{\citenamefont {Ikeda}\ \emph {et~al.}(2019)\citenamefont {Ikeda}, \citenamefont {Straquadine}, \citenamefont {Hristov}, \citenamefont {Worasaran}, \citenamefont {Palmstrom}, \citenamefont {Sorensen}, \citenamefont {Walmsley},\ and\ \citenamefont {Fisher}}]{Ikeda2019}%
  \BibitemOpen
  \bibfield  {author} {\bibinfo {author} {\bibfnamefont {M.~S.}\ \bibnamefont {Ikeda}}, \bibinfo {author} {\bibfnamefont {J.~A.~W.}\ \bibnamefont {Straquadine}}, \bibinfo {author} {\bibfnamefont {A.~T.}\ \bibnamefont {Hristov}}, \bibinfo {author} {\bibfnamefont {T.}~\bibnamefont {Worasaran}}, \bibinfo {author} {\bibfnamefont {J.~C.}\ \bibnamefont {Palmstrom}}, \bibinfo {author} {\bibfnamefont {M.}~\bibnamefont {Sorensen}}, \bibinfo {author} {\bibfnamefont {P.}~\bibnamefont {Walmsley}}, \ and\ \bibinfo {author} {\bibfnamefont {I.~R.}\ \bibnamefont {Fisher}},\ }\bibfield  {title} {\enquote {\bibinfo {title} {{AC} elastocaloric effect as a probe for thermodynamic signatures of continuous phase transitions},}\ }\href {\doibase 10.1063/1.5099924} {\bibfield  {journal} {\bibinfo  {journal} {Review of Scientific Instruments}\ }\textbf {\bibinfo {volume} {90}},\ \bibinfo {pages} {083902} (\bibinfo {year} {2019})},\ \Eprint
  {http://arxiv.org/abs/https://pubs.aip.org/aip/rsi/article-pdf/doi/10.1063/1.5099924/19779198/083902\_1\_online.pdf} {https://pubs.aip.org/aip/rsi/article-pdf/doi/10.1063/1.5099924/19779198/083902\_1\_online.pdf} \BibitemShut {NoStop}%
\bibitem [{\citenamefont {Li}\ \emph {et~al.}(2022)\citenamefont {Li}, \citenamefont {Garst}, \citenamefont {Schmalian}, \citenamefont {Ghosh}, \citenamefont {Kikugawa}, \citenamefont {Sokolov}, \citenamefont {Hicks}, \citenamefont {Jerzembeck}, \citenamefont {Ikeda}, \citenamefont {Hu}, \citenamefont {Ramshaw}, \citenamefont {Rost}, \citenamefont {Nicklas},\ and\ \citenamefont {Mackenzie}}]{Li2022}%
  \BibitemOpen
  \bibfield  {author} {\bibinfo {author} {\bibfnamefont {Y.-S.}\ \bibnamefont {Li}}, \bibinfo {author} {\bibfnamefont {M.}~\bibnamefont {Garst}}, \bibinfo {author} {\bibfnamefont {J.}~\bibnamefont {Schmalian}}, \bibinfo {author} {\bibfnamefont {S.}~\bibnamefont {Ghosh}}, \bibinfo {author} {\bibfnamefont {N.}~\bibnamefont {Kikugawa}}, \bibinfo {author} {\bibfnamefont {D.~A.}\ \bibnamefont {Sokolov}}, \bibinfo {author} {\bibfnamefont {C.~W.}\ \bibnamefont {Hicks}}, \bibinfo {author} {\bibfnamefont {F.}~\bibnamefont {Jerzembeck}}, \bibinfo {author} {\bibfnamefont {M.~S.}\ \bibnamefont {Ikeda}}, \bibinfo {author} {\bibfnamefont {Z.}~\bibnamefont {Hu}}, \bibinfo {author} {\bibfnamefont {B.~J.}\ \bibnamefont {Ramshaw}}, \bibinfo {author} {\bibfnamefont {A.~W.}\ \bibnamefont {Rost}}, \bibinfo {author} {\bibfnamefont {M.}~\bibnamefont {Nicklas}}, \ and\ \bibinfo {author} {\bibfnamefont {A.~P.}\ \bibnamefont {Mackenzie}},\ }\bibfield  {title} {\enquote {\bibinfo {title} {Elastocaloric determination of the phase
  diagram of {Sr}$_2${RuO}$_4$},}\ }\href {\doibase 10.1038/s41586-022-04820-z} {\bibfield  {journal} {\bibinfo  {journal} {Nature}\ }\textbf {\bibinfo {volume} {607}},\ \bibinfo {pages} {276--280} (\bibinfo {year} {2022})}\BibitemShut {NoStop}%
\bibitem [{\citenamefont {Ikeda}\ \emph {et~al.}(2021)\citenamefont {Ikeda}, \citenamefont {Worasaran}, \citenamefont {Rosenberg}, \citenamefont {Palmstrom}, \citenamefont {Kivelson},\ and\ \citenamefont {Fisher}}]{Ikeda2021}%
  \BibitemOpen
  \bibfield  {author} {\bibinfo {author} {\bibfnamefont {M.~S.}\ \bibnamefont {Ikeda}}, \bibinfo {author} {\bibfnamefont {T.}~\bibnamefont {Worasaran}}, \bibinfo {author} {\bibfnamefont {E.~W.}\ \bibnamefont {Rosenberg}}, \bibinfo {author} {\bibfnamefont {J.~C.}\ \bibnamefont {Palmstrom}}, \bibinfo {author} {\bibfnamefont {S.~A.}\ \bibnamefont {Kivelson}}, \ and\ \bibinfo {author} {\bibfnamefont {I.~R.}\ \bibnamefont {Fisher}},\ }\bibfield  {title} {\enquote {\bibinfo {title} {Elastocaloric signature of nematic fluctuations},}\ }\href {\doibase 10.1073/pnas.2105911118} {\bibfield  {journal} {\bibinfo  {journal} {Proceedings of the National Academy of Sciences}\ }\textbf {\bibinfo {volume} {118}},\ \bibinfo {pages} {e2105911118} (\bibinfo {year} {2021})},\ \Eprint {http://arxiv.org/abs/https://www.pnas.org/doi/pdf/10.1073/pnas.2105911118} {https://www.pnas.org/doi/pdf/10.1073/pnas.2105911118} \BibitemShut {NoStop}%
\end{thebibliography}%

\end{document}